\documentclass[lettersize,journal]{IEEEtran}
\usepackage{amsmath,amsfonts}
\usepackage{algorithmic}
\usepackage{array}
\usepackage[caption=false,font=normalsize,labelfont=sf,textfont=sf]{subfig}
\usepackage{textcomp}
\usepackage{stfloats}
\usepackage{url}
\usepackage{verbatim}
\usepackage{graphicx}
\usepackage{cite}
\graphicspath{{Figures/}{./}}

\usepackage{tabu}                      
\usepackage{booktabs}                  
\usepackage{lipsum}                    
\usepackage{mwe}                       

\usepackage{xcolor}
\usepackage{color, colortbl}
\definecolor{LightGray}{gray}{0.85}
\definecolor{DarkGray}{gray}{0.65}
\definecolor{Black}{gray}{0}
\usepackage[normalem]{ulem}

\newcommand{\revise}[1]{\textcolor{black}{#1}}
\newcommand{\reviseTVCG}[1]{\textcolor{black}{#1}}

\usepackage{mathptmx}                  
\usepackage{amsmath}
\usepackage{amssymb}
\usepackage{xfrac}
\usepackage[linesnumbered,ruled,vlined]{algorithm2e}

\usepackage{fancyvrb,fvextra}

\begin{document}

\title{Multi-Criteria Decision Analysis\\for Aiding Glyph Design}

\author{Hong-Po Hsieh, Amy Zavatsky, and Min Chen
    \IEEEcompsocitemizethanks{
    \IEEEcompsocthanksitem Hong-Po Hsieh, Amy Zavatsky, and Min Chen are with University of Oxford, UK.
    E-mails: hong-po.hsieh@kellogg.ox.ac.uk, \{min.chen, amy.zavatsky\}@eng.ox.ac.uk.}
    \thanks{Manuscript received XXX XX, 2024; revised XXX XX, 2024.}
}


\markboth{VIS2024 Recommended Revision to IEEE TVCG}%
{\MakeLowercase{\textit{H. P. Hsieh et al.}}: Multi-Criteria Decision Analysis for Aiding Glyph Design}


\maketitle
\IEEEdisplaynontitleabstractindextext
\IEEEpeerreviewmaketitle

\begin{abstract}
Glyph-based visualization is one of the main techniques for visualizing complex multivariate data. With small glyphs, data variables are typically encoded with relatively low visual and perceptual precision. Glyph designers have to contemplate the trade-offs in allocating visual channels when there is a large number of data variables. While there are many successful glyph designs in the literature, there is not yet a systematic method for assisting visualization designers to evaluate different design options that feature different types of trade-offs. In this paper, we present an evaluation scheme based on the multi-criteria decision analysis (MCDA) methodology. The scheme provides designers with a structured way to consider their glyph designs from a range of perspectives, while rendering a semi-quantitative template for evaluating different design options. In addition, this work provides guideposts for future empirical research to obtain more quantitative measurements that can be used in MCDA-aided glyph design processes.
\end{abstract}



\section{Introduction}
\label{sec:Intro}

Glyph-based visualization is a family of widely used techniques, which are often integrated with other families of visualization techniques, such as small multiples in geo-spatial visualization, multivariate vertices in network visualization, directional and multivariate feature depiction in volume, vector, and tensor field visualization, and dynamic characteristics of objects in event and video visualization. While there have been proposals and discourses on desirable properties of glyph designs in the literature (e.g., \cite{Bertin:2011:book,Maguire:2012:TVCG,Borgo:2013:STAR,chung2015glyph}), there is not yet a coherent methodology that visualization designers can use, consistently and methodically, in evaluating different design options in a process for designing and developing a visualization solution.

In glyph design processes, a designer may face many challenges (e.g., knowledge about the data, users, tasks, cognitive theories and experimental findings related to glyph-based visualization, and so on). This work focuses on one particular challenge, that is, there are many desirable properties of glyph designs, and likely a good design does not necessarily meet all criteria as one might desire, but embodies a relatively optimized set of trade-offs among the visual representations of different variables. In particular, we propose a methodology for evaluating different design options based on multiple-criteria decision analysis (MCDA) \cite{Ishizaka:2013:book,Azzabi:2020:book}, which is an established and widely adopted methodology in management science for evaluating multiple complementary and conflicting criteria explicitly in decision making. Its applications include business, governance, medicine, and engineering. 

We aim to introduce MCDA to glyph design as a systematic and cost-effective methodology and to bring together different desirable properties proposed in the literature into a typology of rateable criteria.
Attention has been paid to (1) providing a good coverage of all proposed criteria for static glyph designs, (2) defining each criterion to facilitate clear interpretation and unambiguous ratings, (3) reducing the overlapping among criteria, (4) enabling distinct considerations of conflicting criteria, and (5) recommending a weighted scoring mechanism that balances between overview vs. detail and precision vs. cost.

\revise{The proposed methodology is built on a comprehensive study of the literature on glyph-based visualization as summarized in Section \ref{sec:RelatedWork}. In particular, the 12 criteria in the proposed MCDA scheme are carefully selected following an in-depth analysis of four existing sets of criteria in the literature \cite{Bertin:2011:book,Maguire:2012:TVCG,Borgo:2013:STAR,chung2015glyph}, which is detailed in Appendix \ref{apx:FrameworkDesign} in the supplementary materials. The analysis enabled us to identify the overlapping or partially overlapping criteria as well as a few missing criteria, while allowing us to gain an appreciation about the scientific reasons behind their categorical arrangement of different criteria. The main contributions of this work include:
\begin{itemize}
    \item The introduction of the MCDA methodology in glyph design processes for evaluating glyph designs systematically;
    \item The integration of four existing sets of criteria for evaluating glyph designs, and the identification and addition of a few criteria absent in the existing sets;%
    \vspace{-1mm}
    \item The definition of five scales for rating each of the 12 criteria, featuring quantitative and meticulous specifications;%
    \item The testing and demonstration of the proposed MCDA scheme through \reviseTVCG{scoring examples} where the scheme was applied to several existing glyph designs in the literature and their parody versions as well as several new designs in a biomechanical application.%
\end{itemize}}

The proposed methodology represents a major step forward from the four existing sets of criteria. The definitions in the scheme are expected to be improved through the experience gained in its uses in practice as well as new findings in theoretical and empirical research.
Furthermore, the proposed methodology is not a replacement for user-centered requirements analysis and evaluation, but it can complement, strengthen, and reduce the frequencies of user-centered studies in individual design processes.
We will discuss these further in Section \ref{sec:Conclusions}.
%

\section{Related Work}
\label{sec:RelatedWork}
Glyph visualization has been applied in many fields, such as
biology and medicine \cite{Maguire:2012:TVCG,somarakis2019imacyte,meuschke2017glyph,raidou2018bladder,lichtenberg2017concentric,oeltze2008glyph,meyer2008glyph},
meteorology and environmental studies \cite{martin2008results,pilar2013representing,pfeiffer2021glyph,drocourt2011temporal,sanyal2010noodles},
human behavioral analysis \cite{el2016contovi,kovacevic2020glyph},
web and database searching \cite{chau2011visualizing,siva2014evaluation},
sports \cite{legg2016glyph,polk2014tennivis,wang2021tac,wu2021tacticflow,cava2013glyphs,parry2011hierarchical},
music and multimedia \cite{chan2009visualizing,lind2022visualizing,janicke2010soundriver,botchen2008action},
and business and industrial applications \cite{rees2020agentvis,surtola2005effect,suntinger2008event}.
A glyph object conveys a multivariate data record in a concise way, which significantly reduces the perceptual and cognitive load for information comprehension. It has been widely adopted in applications that require a simultaneous view of multiple variables and multiple data records, including
the facilitation of visual search \cite{healey1999large,cai2015applying},
data comparison \cite{verma2004comparative,zhang2015glyph,meuschke2017glyph,koc2022peaglyph},
data ordering \cite{Chung:2016:CGF,miller2019evaluating},
and feature extraction \cite{keck2017towards}.
Glyphs have also been used in combination with spatial or temporal visualization \cite{tominski20053d,ropinski2007surface,drocourt2011temporal,bleisch2017exploring,Legg:2012:CGF}
and other visualization techniques \cite{lichtenberg2017concentric,kammer2020glyphboard,fernstad2020explore,borgo2012empirical}.
Algorithms for glyph placement \cite{ward2002taxonomy,lie2009critical,streeb2018design,rees2020agentvis,tong2016glyphlens,mcnabb2019multivariate,hlawitschka2007interactive}
and 3D visualization \cite{lie2009critical,tong2016glyphlens,stevens2016hairy} were developed for presenting glyph objects. Techniques for specialized scenarios or requirements include
temporal summarization \cite{Duffy:2015:TVCG,el2016contovi,gerrits2017glyphs,tominski20053d,botchen2008action},
uncertainty depiction \cite{aigner2005planninglines,sanyal2010noodles,hlawatsch2011flow,wittenbrink1996glyphs},
visualizing data with special structures or inter-relations \cite{rees2020agentvis,dunne2013motif,cayli2013glyphlink,lee2021cluster,soares2020depicting,cao2011dicon,reda2019dynamic},
and depiction of directional and high-dimensional information in vector and tensor fields (e.g., \cite{schultz2010superquadric,gerrits2016glyphs,tong2016crystal,meuschke2017glyph,hergl2019visualization})
Attempts were made to develop tools for glyph generation \cite{ribarsky1994glyphmaker,xia2018dataink,brehmer2021generative,ying2022metaglyph,cunha2018many}.

With a huge design space and the wide variety of applications, it is highly desirable to provide glyph designers with guidance and methods for evaluating different designs.
Empirical studies have been used to evaluate glyph designs \cite{aigner2005planninglines,surtola2005effect,weigle2005visualizing,chan2009visualizing,chau2011visualizing,siva2014evaluation,dunne2013motif}.
While empirical studies are useful for gauging design qualities from user experience and feedback, they incur arduous time and effort and usually cannot be conducted frequently during a design process.   
A survey conducted by Fuchs et al. \cite{fuchs2016systematic} revealed that two task performance measures (accuracy and completion times) were widely adopted for evaluating the effectiveness of glyph designs.

Normative rating is another way to evaluate visualizations, in which design qualities can be individually quantified. User-centered subjective rating is commonly used in empirical studies \cite{lee2003empirical,aigner2005planninglines,surtola2005effect,fuchs2013evaluation}. The rating schemes used in empirical studies normally do not require design expertise. Normative rating led by visualization experts has not been widely reported in the field of visualization. In psychology, McDougall et al. \cite{mcdougall2000exploring} adopted subjective ratings to characterize cognitive features of icon designs and used the ratings to investigate the correlations among different design qualities identified by experts of icon designers.

Attempts have been made to develop standardized measures for qualities of glyph visualization. Garcia et al. \cite{garcia1994development} proposed a metric to evaluate glyph complexity.
Forsythe et al. \cite{forsythe2003measuring} proposed an automatic complexity measuring algorithm, aiming to remove subjective elements from the evaluation of complexity and to  prevent judgment bias. Up to now, however, the number of standardized measures of visual design qualities remains very limited.

A survey conducted by Borgo et al. \cite{Borgo:2013:STAR} collected a good number of glyph design guidelines and criteria in the literature. They cover different levels of glyph design (e.g., variable encoding, inter-channel interaction, and holistic glyph design) and different aspects of glyph-based visualization (e.g., 2D and 3D designs, interaction, and placement).
At the variable encoding level, Bertin \cite{bertin1983semiology} proposed four basic perceptual criteria for channel encoding. Cleveland and McGill \cite{cleveland1984graphical}, Mackinlay \cite{Mackinlay:1986:TOG}, and Munzner \cite{Munzner:2014:book} recommended the ordering of visual channels. Other considerations include the capacity, orderability, semantic closeness, visual pre-attentiveness, robustness, and normalizability of the channels \cite{lie2009critical,Legg:2012:CGF,ropinski2011survey,ward2008multivariate,chung2015glyph}.
At the levels of inter-channel interactions and holistic glyph design, guidelines were proposed for the integration and separability of channels, the balance of attention, searchability, visual hierarchy, and the learnability of glyph designs \cite{karve2007glyph,Maguire:2012:TVCG,lie2009critical,chung2015glyph}. Beyond the design of the glyph object \textit{per se}, advice was provided for other aspects of data mapping and glyph rendering \cite{ropinski2007surface,ward2008multivariate,meyer2008glyph,ropinski2008taxonomy,lie2009critical,ropinski2011survey}.
All these proposed guidelines and evaluation criteria prepared us for developing normative rating methods to be used by visualization experts. Such methods could be employed frequently in a design process, complementing user-centered empirical studies.

\section{Overview, Terminology, and Design Considerations}
\label{sec:Overview}
\reviseTVCG{As mentioned in Section \ref{sec:RelatedWork}, a number of criteria for glyph design have been proposed in the literature \cite{Bertin:2011:book,Maguire:2012:TVCG,Borgo:2013:STAR,chung2015glyph}. After a comparative study of these criteria, we regrouped them into 12 criteria that will be detailed in Section \ref{sec:Scheme}. The rationale of the regrouping is discussed in Appendix \ref{apx:FrameworkDesign}.} 

In this work, we adopt the narrow definition of \emph{glyph} given by Borgo et al. \cite{Borgo:2013:STAR}, i.e., ``a glyph is a small independent visual object that depicts
attributes of a data record; glyphs are discretely placed in a display space; and
glyphs are a type of visual sign but differ from other types of signs such as icons, indices and symbols.''
We focus only on 2D glyphs.
Consider a multivariate data record for storing the values of $m$ \emph{variables}, $D = \{d_1, d_2, \ldots, d_m\}$, and a glyph for encoding such a record using $n$ \emph{visual channels} $\Gamma = \{\gamma_1, \gamma_2, \ldots, \gamma_n\}$. Since each data variable can be encoded using multiple visual channels, we have $n \geq m$.

Multiple-criteria decision analysis (MCDA) \cite{Ishizaka:2013:book,Azzabi:2020:book} can be viewed as a tree-based scoring system, where the score of each node is a weighted average of the scores at its child-nodes. Here an unweighted average is considered as a special case. There are a few high-level design considerations, e.g., do we decompose the holistic assessment into the assessments of data variables or visual channels, and how detailed do the assessments need to be?     

\vspace{1mm}
\noindent\textbf{Design Consideration 1. Data Variables vs Visual Channels.} 
In the previous works on glyph design, some criteria are defined for accessing visual channels (e.g., channel capacity \cite{chung2015glyph}), while others are applicable to data variables (e.g., metaphoric representation \cite{Maguire:2012:TVCG}).
When one assumes a 1-to-1 mapping between a data variable and a visual channel, evaluating a data variable implicitly implies the evaluation of the corresponding visual channel, and vice versa. However, when a data variable is encoded using multiple visual channels, assessing individual channels independently may not inform their combined effect on perception and interpretation of the data variable concerned.
Hence the scheme places an emphasis on data variables. 

\vspace{1mm}
\noindent\textbf{Design Consideration 2. Hierarchy of MCDA Evaluation.}
The number of criteria may decrease (or increase), if such criteria are combined (or finely decomposed) in evaluation.
Some criteria may be applicable to each data variable or visual channel (e.g., Bertin's criteria \cite{Bertin:2011:book}), while others may involve assessing a group of data variables (e.g., separability \cite{chung2015glyph}), or a whole glyph (e.g., attention balance \cite{chung2015glyph}). There are four types of basic assessment modes:
\begin{itemize}
    \item \textbf{Type A.} A criterion is for evaluating the visual encoding of a data variable $d_i$ independently. Given $m$ data variables, there are $m$ assessments for each Type A criterion.%
    \item \textbf{Type B.} A criterion is for evaluating how the encoding of a data variable $d_i$ is affected by the encoding of another $d_j (j \neq i)$. Given $m$ variables, there are potentially $m \times (m-1)$ assessments as the impact of $d_i$ on $d_j$ may not be the same as the impact of $d_j$ on $d_i$.%
    \item \textbf{Type C.} A criterion is for evaluating how the encoding of a data variable $d_i$ is affected by that of all others, i.e., $\forall d_j \in \{d_1, d_2, \ldots, d_m\} (j \neq i)$. There are $m$ assessments for $m$ variables.%
    \item \textbf{Type D.} A criterion is for evaluating the whole glyph holistically. There is one assessment per glyph for each Type D criterion.
\end{itemize}

\begin{figure}[t!]
    \centering
    \includegraphics[width=\linewidth]{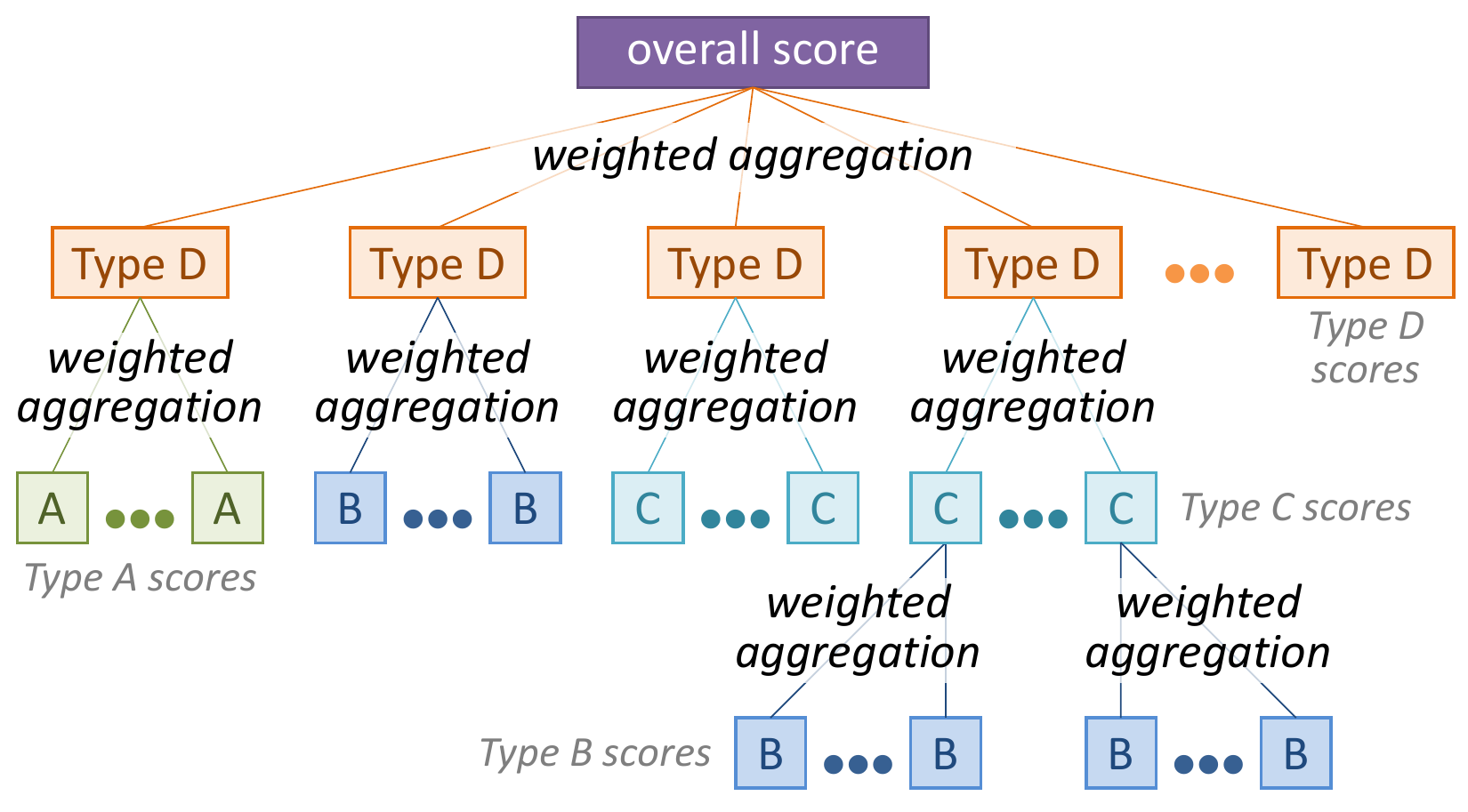}
    \caption{Weighted aggregation is commonly used in hierarchical MCDA. For example, a Type D criterion may be scored directly, or assessed by aggregating the scores of Type A, Type B, or Type C criteria.}
    \label{fig:Hierarchy}
\end{figure}

As illustrated in Fig.~\ref{fig:Hierarchy}, the score of a high-level criterion (e.g., Type D) can often be obtained from scores of low-level criteria (e.g., Type A and Type B) through aggregation.
If a MCDA process consists of many low-level criteria, the designer will need to determine scores for individual variables or pairs of variables. In comparison with producing a high-level score directly (i.e., not through aggregation), the low-level scores are usually more precise, easier to judge, but more time-consuming to obtain.
Hence, there is a trade-off between precision and time cost.
In our proposed MCDA scheme, we indicate the possible modes, in which each criterion could be assessed, and recommend a specific mode of assessment for optimizing the trade-off concerned.   

\vspace{1mm}
\noindent\textbf{Design Consideration 3. Multiple complementary or conflicting facets.}
In the literature, a suggested criterion may consist of two or more facets. For example, attention balance \cite{chung2015glyph} encourages glyph designers to allow the encoding of important data variables to attract more attention, while making sure other data variables are not seriously disadvantaged. To ease the assessment, we intentionally divide such a multi-facet criterion into two criteria at the same level, allowing the designers to score the two potentially conflicting criteria independently.

\section{An MCDA Scheme for Glyph Design}
\label{sec:Scheme}
Our proposed scheme consists of 12 criteria described in the following subsections (see also Appendices \ref{apx:FrameworkDesign} and \ref{apx:Workflow}). For each criterion, we provide a definition, a specification of five-level scores [1$\sim$5], a recommended weight, and a recommended mode of assessment (in black) and other possible modes (in grey).  

\subsection{Typedness (Criterion 1)}
\label{sec:Typedness}
\begin{itemize}
\item \textbf{Definition:} This criterion assesses whether or not the visual channel (channels) of a data variable is (are) appropriately selected to match the data type of the variable to be encoded. Such data types may include, but are not limited to: \emph{nominal}, \emph{ordinal}, \emph{interval}, \emph{ratio}, and \emph{directional}.%
\item \textbf{Recommended Modes:} Type A (direct -- \emph{directly assessed}), Type D (aggregated -- \emph{using Type A scores}).%
\item \textbf{Recommended Weight:} Type A (unweighted -- \emph{all have weight 1 when being aggregated}), Type D [1.0].
\end{itemize}

The above definition was proposed by Chung et al. \cite{chung2015glyph,Borgo:2013:STAR} for assessing each visual channel in a glyph based on Bertin's four kinds of perception (KOP), i.e., \emph{associative}, \emph{selective}, \emph{ordered}, and \emph{quantitative} perception.
Not all KOPs are applicable to all data types. For example, \emph{ordered} and \emph{quantitative} perception are not both required for a nominal variable. Hence, we consider only applicable KOPs (i.e., AKOPs in short). See also Appendices \ref{apx:Bertin} and \ref{apx:Typedness} in the supplementary materials.%

We consider three levels of typedness, \emph{appropriate}, \emph{usable}, and \emph{inappropriate}. 
When a data variable $d_i$ is encoded using  just one visual channel $\gamma_i$, we may derive a score $s_i$ as follows:    
\begin{enumerate}
    \item[5.] The visual channel is appropriate for all AKOPs.
    \item[4.] The visual channel is appropriate for some AKOPs and is usable for the other AKOPs.
    \item[3.] The visual channel is usable for all AKOPs.
    \item[2.] The visual channel is inappropriate for some AKOPs.
    \item[1.] The visual channel is inappropriate for all AKOPs.
\end{enumerate}

When a data variable $d_i$ is encoded using two or more visual channels $\gamma_{i,1}, \gamma_{i,2}, \ldots$, we consider the best visual channel for each KOP. Hence the above scores are redefined as:

\begin{enumerate}
    \item[5.] For each AKOP, at least one visual channel is appropriate.
    \item[4.] For some AKOPs, the best visual channel is appropriate, while for the other AKOPs, the best is usable.
    \item[3.] For each AKOP, the best visual channel is usable.
    \item[2.] For some AKOPs, all visual channels are inappropriate.
    \item[1.] For each AKOP, all visual channels are inappropriate.
\end{enumerate}

\subsection{Discernability (Criterion 2)}
\label{sec:Discernability}
\begin{itemize}
\item \textbf{Definition:} This criterion assesses whether the encoding of a data variable allows viewers to differentiate key values or major value ranges. The encoding may use one or more visual channels.%
\item \textbf{Recommended Modes:} Type A (direct), Type D (aggregated).%
\item \textbf{Recommended Weight:} Type A (unweighted), Type D [1.0].
\end{itemize}

A data variable may have a number of valid values in an application context. For some numerical variables, the number of values can be huge. Glyph-based visualization is normally not intended for users to observe such data variables at a high-resolution (see Appendix \ref{apx:Discernability}).
It is therefore helpful to define a set of key values (or key data ranges) for each data variable in an application context.  
Given $k$ key values (or key value ranges), viewers potentially need to differentiate $n = k(k-1)/2$ pairs of values (ranges). We define five levels based on $n$.

\begin{enumerate}
    \item[5.] All $n$ pairs of values (ranges) can be differentiated at ease.%
    \item[4.] Most of the $n$ pairs of values (ranges) can be differentiated at ease, and the rest are perceptually differentiable. Numerically, ``most'' is defined as $[75, 100)$, i.e., $\geq 75\%$ and $< 100\%$.%
    \item[3.] A large portion of the $n$ pairs of values (ranges) can be differentiated at ease, and the rest are perceptually differentiable. Numerically, ``large portion'' is defined as $[50, 75)$, i.e., $\geq 50\%, < 75\%$.%
    \item[2.] Most of the $n$ pairs of values (ranges) are perceptually differentiable. The rest are not. As for [4.], ``most'' is defined as $[75, 100)$.%
    \item[1.] A significant portion of the $n$ pairs of values (ranges) are not perceptually differentiable. Numerically, ``significant portion'' is defined as [25, 100], i.e., $\geq 25\%$.
\end{enumerate}

\begin{figure}[t!]
  \centering
  \includegraphics[width=\linewidth]{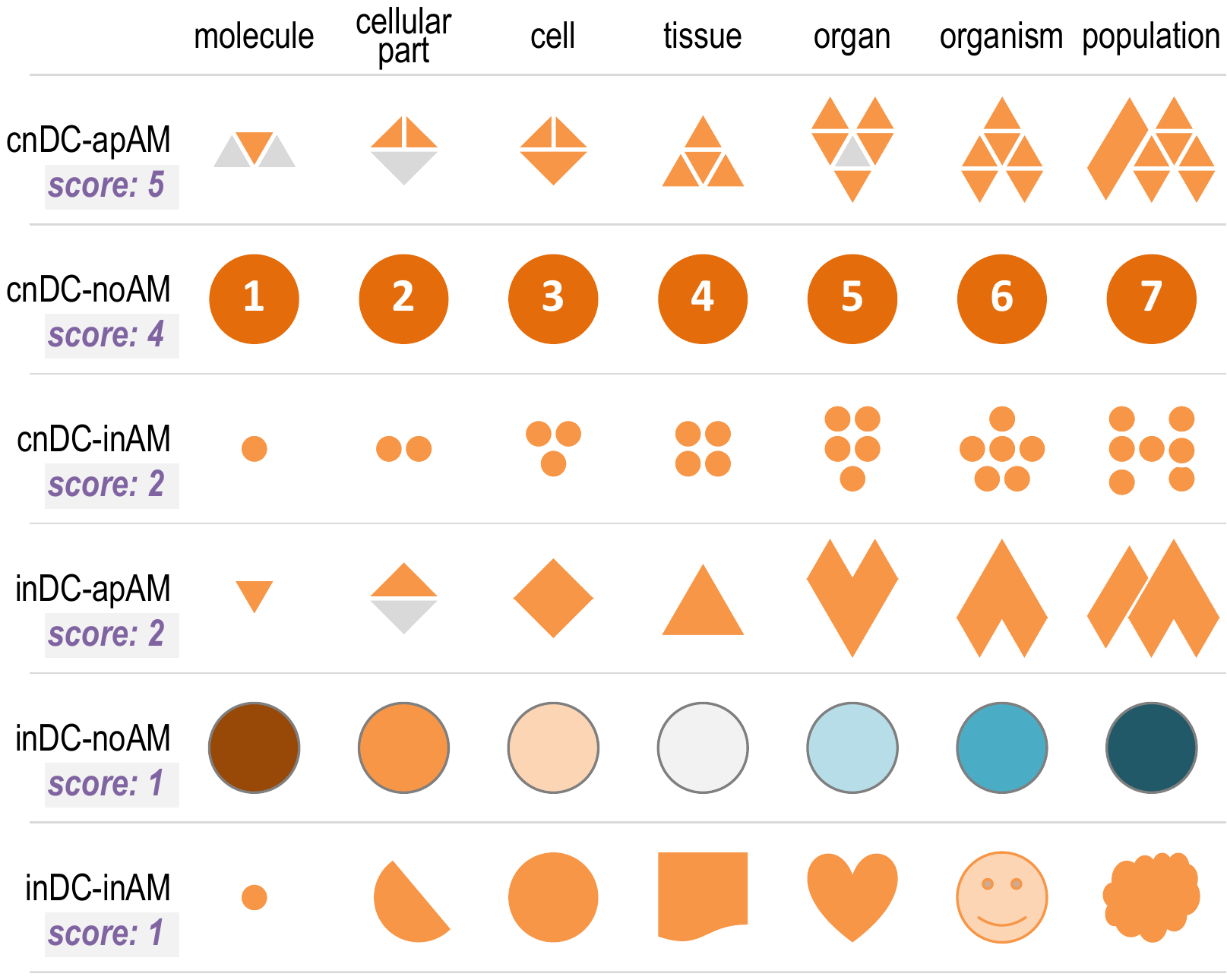}
  \caption{Maguire et al. \cite{Maguire:2012:TVCG} discussed the encoding of seven levels of material granularity in a biological application. The intuitiveness of six encoding methods is assessed here.}
  \label{fig:Intuitiveness}
\end{figure}

\subsection{Intuitiveness (Criterion 3)}
\label{sec:Intuitiveness}
\begin{itemize}
\item \textbf{Definition:} This criterion assesses how the encoding of a data variable is semantically-related to the knowledge of viewers, and how such a relation makes the encoding knowable to the viewers by intuition.%
\item \textbf{Recommended Modes:} Type A (direct), Type D (aggregated).%
\item \textbf{Recommended Weight:} Type A (unweighted), Type D [1.0].
\end{itemize}

Given a data variable $d_i$ in an application domain, there may be a \emph{Domain-specific Convention} (DC) for encoding the data variable visually. Such encoding captures the semantic knowledge of the discipline, hence facilitating intuitiveness. Given a design option for encoding a data variable $d_i$, there are three basic scenarios of DC: no existing DC (noDC), consistent with the existing DC (cnDC), and inconsistent with the existing DC (inDC).

The encoding may also introduce one or more \emph{Additional Visual Metaphors} (AM) for enhancing the existing DC or filling the gap when there is no existing DC. An appropriate metaphor can make intuitive connection between the semantics of $d_i$ and viewers' knowledge. Hence, we consider four basic scenarios of AM: no additional metaphor (noAM), appropriate metaphor (apAM), adequate metaphor (okAM), and inappropriate metaphor (inAM). 
The two sets of scenarios result in 12 combinations. We define five levels as:
\begin{enumerate}
    \item[5.] cnDC-apAM or noDC-apAM.
    \item[4.] cnDC-noAM, cnDC-okAM, or noDC-okAM.
    \item[3.] noDC-noAM.
    \item[2.] cnDC-inAM, inDC-okAM, inDC-apAM.
    \item[1.] noDC-inAM, inDC-noAM, inDC-inAM.
\end{enumerate}
\noindent Here we do not assume that DC is always better than AM, or vice versa. Nevertheless, the criterion awards the creativity of visual design by rating noDC-apAM slightly higher than cnDC-noAM. Examples of how this criterion can be applied are shown in Fig.~\ref{fig:Intuitiveness}. See also Appendix \ref{apx:Intuitiveness}.

\subsection{Invariance: Geometry (Criterion 4)}
\label{sec:Geometry}
\begin{itemize}
\item \textbf{Definition:} This criterion assesses the undesirable impact of the geometrical variations in displaying a glyph upon its visual quality. The primary geometrical variations are size variations. Other considerations include minor variations of aspect radio, projection angle, and rotation.%
\item \textbf{Recommended Modes:} \textcolor{gray}{Type A,} Type D (direct).%
\item \textbf{Recommended Weight:} Type D [0.5].
\end{itemize}

Maguire et al. \cite{Maguire:2012:TVCG} conducted scalability tests in their glyph design process. Legg et al. \cite{legg2016glyph} studied the geometric and color degeneration of glyphs. To reduce the complexity of assessing these two types of degeneration (i.e., Design Consideration 2), we have two sub-criteria, geometry (this section) and colorimetry (next section), under the common quality term ``Invariance''.

\begin{figure}[t!]
    \centering
    \includegraphics[width=\linewidth]{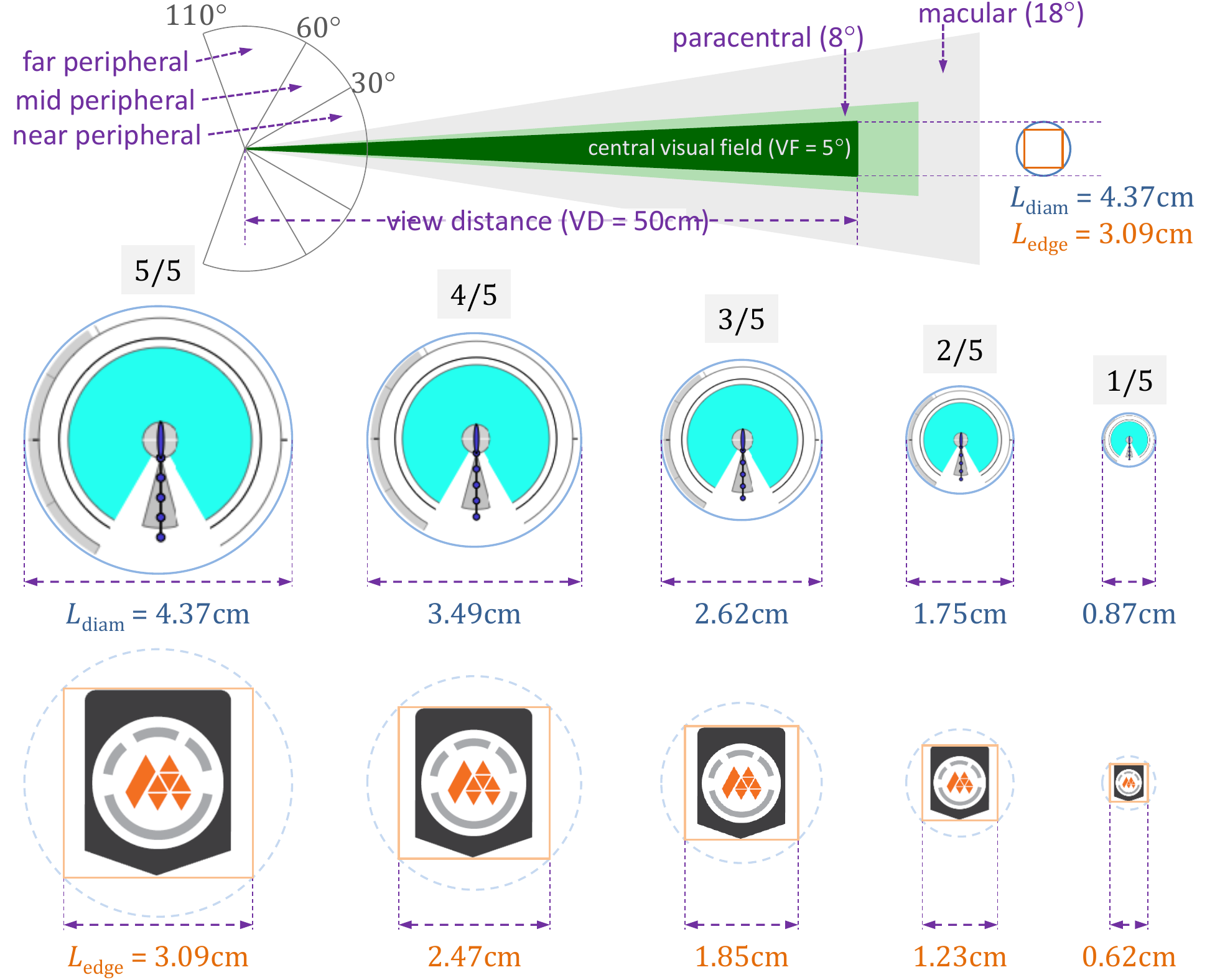}
    \caption{The view area depends on two main factors, visual field and viewing distance. Five scaling factors are applied to a circular glyph \cite{Duffy:2015:TVCG} and a rectangular glyph \cite{Maguire:2012:TVCG}.}
    \label{fig:Invariance}
\end{figure}

In vision science and human-computer interaction, there is a wealth of research on viewing distances and icon sizes. We can build on these findings to consider the geometric scalability of glyphs. As illustrated in Fig.~\ref{fig:Invariance}, the viewing area can be approximated by a cone that has a visual field VF degrees and a viewing distance VD. The central VF of human eyes is $1.5^\circ \sim 5^\circ$ \cite{wandell1995foundations}. Typical viewing distances for monitors, laptops, and phones are  $50\sim100$~cm, $40\sim76$~cm, and $41\sim46$~cm, respectively. We can estimate the viewing area as a circle with diameter $L_\text{diam}$ or its inner square with edge length $L_\text{edge}$ as follows:
%
\[
    L_\text{diam} = 2 \cdot \text{\text{VD}} \cdot \tan(\text{VF}^\circ), \qquad L_\text{edge} = L_\text{diam} / \sqrt{2}
\]
\noindent In this work, we use $\text{VF}=5^\circ$ \cite{wandell1995foundations,strasburger2011peripheral} and $\text{VD}=50$~cm (widely-recommended viewing distance) as the baseline measures. The baseline of a circular viewing area has a diameter of 4.37~cm and a square with an edge length of 3.09~cm as illustrated in Fig.~\ref{fig:Invariance}.
We can define the five levels based on scaling factors $1/5,\,2/5,\, 3/5,\, 4/5,\, 5/5$ and the absence of any degradation of discernability of any visual channel (i.e., invariance) at each scale.  
\begin{enumerate}
    \item[5.] Discernability is invariant at the $1/5$ scale.%
    \item[4.] Discernability is invariant at the $2/5$ scale but variant at $1/5$.%
    \item[3.] Discernability is invariant at the $3/5$ scale but variant at $2/5$.%
    \item[2.] Discernability is invariant at the $4/5$ scale but variant at $3/5$.%
    \item[1.] Discernability is variant at the $4/5$ scale.
\end{enumerate}
As exemplified by Maguire et al. \cite{Maguire:2012:TVCG}, such invariance tests can easily be carried out by glyph designers. Since one will likely apply a scaling factor to a whole glyph, we recommend to assess this criterion with a Type D score directly. See also Appendix \ref{apx:Geometry}. 

\subsection{Invariance: Colorimetry (Criterion 5)}
\label{sec:Colorimetry}
\begin{itemize}
\item \textbf{Definition:} This criterion assesses the undesirable impact of the non-geometrical appearance variations in displaying a glyph upon its visual quality. The main considerations are chromatic and achromatic variations of colors, which may be caused by limitations of a display device and/or environmental lighting conditions.%
\item \textbf{Recommended Modes:} \textcolor{gray}{Type A,} Type D (direct).%
\item \textbf{Recommended Weight:} Type D [0.5].
\end{itemize}

Chromatic and achromatic variations of colors are commonly caused by undesirable environmental lighting (e.g., reflection) and occasionally by display devices (e.g., in an energy saving mode). Such variations can be approximated by using a function for varying the contrast and brightness of the imagery representation of a glyph. Consider the RGB representation of a pixel, such that $r, g, b \in [0, 255]$. One commonly-used function \cite{Loch:2021:web} is:
%
\[
    x' = \min\biggl(255, \max\bigl(0, \frac{259(\kappa_\text{ctr} + 255)}{255(259-\kappa_\text{ctr})} (x - 128) + 128 + \kappa_\text{brt} \bigr) \biggr)
\]
\noindent where $\kappa_\text{ctr} \in [-255, 255]$ specifies the increment (positive) or decrement (negative) of contrast, while $\kappa_\text{brt} \in [-255, 255]$ specifies the increment or decrement of brightness. We recommend to assess this criterion with a Type D score. We define the five levels based on 10\%, 20\%, 30\%, and 40\% of variations, which are translated to $\pm 25.5, \pm 51, \pm 76.5,$ and $\pm 102$ for $\kappa_\text{ctr}$ and $\kappa_\text{brt}$. See also Appendix \ref{apx:Colorimetry}.

\begin{enumerate}
    \item[5.] Discernability is invariant with $\kappa_\text{ctr} = \pm 102 \land \kappa_\text{brt} = \pm 102$.%
    \item[4.] Discernability is invariant when $\kappa_\text{ctr} = \pm 76.5 \land \kappa_\text{brt} = \pm 76.5$, but variant when $\kappa_\text{ctr} = \pm 102 \lor \kappa_\text{brt} = \pm 102$.%
    \item[3.] Discernability is invariant when $\kappa_\text{ctr} = \pm 51 \land \kappa_\text{brt} = \pm 51$, but variant when $\kappa_\text{ctr} = \pm 76.5 \lor \kappa_\text{brt} = \pm 76.5$.%
    \item[2.] Discernability is invariant when $\kappa_\text{ctr} = \pm 25.5 \land \kappa_\text{brt} = \pm 25.5$, but variant when $\kappa_\text{ctr} = \pm 51 \lor \kappa_\text{brt} = \pm 51$.%
    \item[1.] Discernability is variant when $\kappa_\text{ctr} = \pm 25.5 \lor \kappa_\text{brt} = \pm 25.5$.%
\end{enumerate}

\revise{Similar to the criterion of geometry invariance in Section \ref{sec:Geometry}, we recommend to assess this criterion with a Type D score directly because one will likely test the invariance of colorimetry by manipulating a whole glyph.}

\subsection{Composition: Separability ((Criterion 6)}
\label{sec:Separability}
\begin{itemize}
\item \textbf{Definition:} This criterion assesses the undesirable interference among visual channels in a glyph, which would affect the perception of some visual channels.%
\item \textbf{Recommended Modes:} \textcolor{gray}{Type B, Type C,} Type D (direct).%
\item \textbf{Recommended Weight:} Type D [0.5].
\end{itemize}

\begin{figure*}[ht]
    \centering
    \includegraphics[width=180mm]{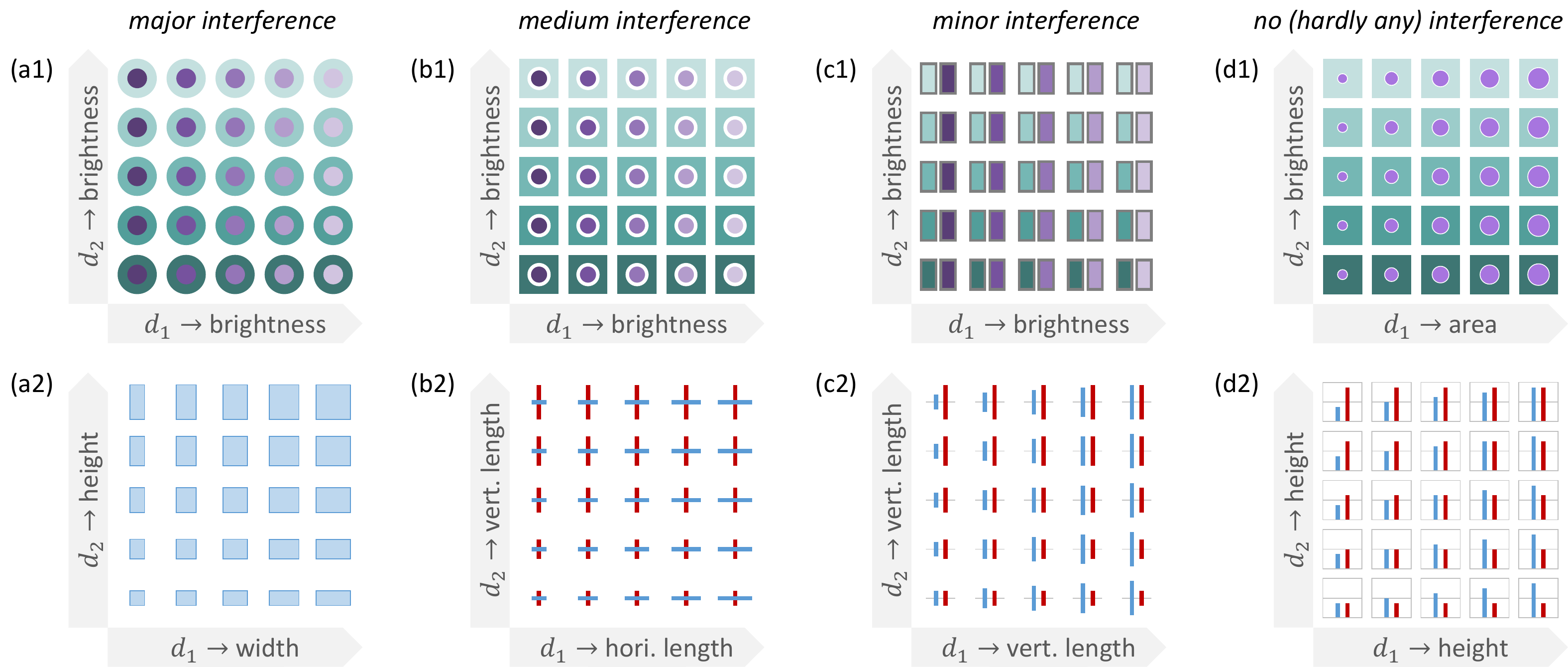}
    \caption{Example designs of bi-variate glyphs in four severity categories. We can observe that introducing boundaries to separate different visual channels (b1-d1) or reference lines (c2, d2) can reduce the severity of the interference between two visual channels.}
    \label{fig:Separability}
\end{figure*}

This criterion is based on the discourse by Maguire et al. \cite{Maguire:2012:TVCG} and Chung et al. \cite{chung2015glyph} on the interference among visual channels in glyph design. Such interference is often caused by visual channels that are integrated into the same visual object or hosted by closely-placed visual objects. Visual channels of similar types (e.g., brightness, length) are more likely to suffer from interference as exemplified in Fig.~\ref{fig:Separability}. 
In general, a glyph occupies a small display space, and interference is often unavoidable.
Such minor interference would normally be considered to be acceptable in glyph design. In addition, for some applications, it is desirable to make some visual channels comparable, which conflicts with the desire for separating visual channels by using different types of visual channels or moving them away from each other. To balance these two sides of the same ``Composition'' coin, we have introduced ``comparability'' as a distinct criterion.

Although many empirical studies have evaluated the interference among different visual channels, there is not yet a standard metric for measuring the severity of such interference. While such a standard metric will hopefully be defined in future research, we hereby use a relatively subjective measure to categorize interference in glyph design as \emph{major}, \emph{medium}, \emph{minor}, and \emph{none}, which correspond to interference scores 1, 0.1, 0.01, and 0 respectively. Fig.~\ref{fig:Separability} shows two examples for each severity category.  

Given a total of $n$ visual channels $\lambda_1, \lambda_2, \ldots \lambda_n$ in a glyph, each channel $\lambda_i$ can potentially be influenced by $n-1$ other channels. Let $s_\text{int}(\lambda_i, \lambda_j)$ be the score of the interference received by $\lambda_i$ from $\lambda_j$, and $S_\text{int}(\lambda_i)$ be the aggregated score for $\lambda_i$, which is defined as:
%
\[
    S_\text{int}(\lambda_i) = \max \bigl\{
    s_\text{int}(\lambda_i, \lambda_j) \; | \; j=1,2, \ldots, n \;\land\; j \neq i \bigr\}
\]
\noindent $s_\text{int}(\lambda_i, \lambda_j)$ is a pairwise score for Type B assessment, and $S_\text{int}(\lambda_i)$ is a score for Type C assessment.
From these scores, we can obtain two Type D scores, a mean score $\text{avg}_\text{int}$ and a maximum score $\max_\text{int}$:
%
\[
    \text{avg}_\text{int} = \frac{1}{n} \sum_{i=1}^n S_\text{int}(\lambda_i) \qquad
    \text{max}_\text{int} = \max\limits_{i=1}^n S_\text{int}(\lambda_i)
\]
\noindent We can therefore define the five levels as:
\begin{enumerate}
    \item[5.] $0.0 \leq \max_\text{int} < 0.1$: only minor interference.%
    \item[4.] $0.1 \leq \max_\text{int} < 1.0$: some medium but no major interference.%
    \item[3.] $\max_\text{int} = 1.0 \;\land\; \text{avg}_\text{int} < 1/8$: some major interference, and less than $1/8$ of visual channels are affected.
    \item[2.] $\max_\text{int} = 1.0 \;\land\; 1/8 \leq \text{avg}_\text{int} < 1/4$: some major interference, and between $1/8$ and $1/4$ of visual channels are affected.
    \item[1.] $\max_\text{int} = 1.0 \;\land\; \text{avg}_\text{int} \geq 1/4$: some major interference, and more than $1/4$ of visual channels are affected.
\end{enumerate}
\noindent As obtaining $n(n-1)$ pairwise scores $s_\text{int}(\lambda_i, \lambda_j)$ will be time-consuming, we recommend to evaluate each glyph design holistically. The above specification of the five levels facilities the option of obtaining a Type D score directly. See also Appendix \ref{apx:Separability}.

\subsection{Composition: Comparability (Criterion 7)}
\label{sec:Comparability}
\begin{itemize}
\item \textbf{Definition:} This criterion assesses the desirable level of support featured in a glyph design for enabling required comparative tasks such as determining the order of two related data variables ($d_i$ vs. $d_j$), estimating their addition $(d_i+d_j)$, their difference $|d_i - d_j|$, or their ratio $d_i / d_j$.%
\item \textbf{Recommended Modes:} \textcolor{gray}{Type B, Type C,} Type D (direct).%
\item \textbf{Recommended Weight:} Type D [0.5].
\end{itemize}

As mentioned in Section \ref{sec:Separability}, this criterion complements ``Separability''. It was not discussed explicitly in the existing surveys (e.g., \cite{Borgo:2013:STAR}), possibly because one only considers this when there is a need to compare some data variables within a glyph. Nevertheless, several glyph designs in the literature addressed the need to compare some data variables within a glyph representation. For example, Duffy et al. \cite{Duffy:2015:TVCG} presented a glyph design representing some 20 data variables, among which three related distance variables were to be compared in terms of their lengths and relative ratios. Duffy et al. encoded these variables using three nested arcs, facilitating easy comparison.
When one needs to compare two visual channels that encode two different data variables, the following obstacles will likely hinder comparative tasks:
\begin{itemize}
    \item \emph{Major obstacle} -- Two visual channels are of very different types, e.g., area vs. brightness as shown in Fig.~\ref{fig:Separability} (d1).%
    \item \emph{Major obstacle} -- Two visual channels are of the same type but with inconsistent encoding schemes, e.g., two color channels, one with a divergence colormap and another with a sequential colormap, or two length channels, one uses 20 pixels to encode the range [0, 10] and another uses 40 pixels for the same range.%
    \item \emph{Medium obstacle} -- Two visual channels are of the same type but with features that affect consistent perception, e.g., same brightness range but with different hues in Fig.~\ref{fig:Separability} (c1), or same length encoding but in different orientation in Fig.~\ref{fig:Separability} (b1).%
    \item \emph{Medium obstacle} -- Two visual channels do not have any common reference point, e.g., two length channels without any reference lines, unlike Fig.~\ref{fig:Separability} (c2, d2).%
    \item \emph{Minor obstacle} -- Two visual channels are placed far away from each other, where the word ``far'' is in the context of a glyph. 
\end{itemize}
 
It is necessary to note that, when a data variable is encoded using multiple visual channels, as long as one of the channels is comparable, one may omit the consideration of other channels. For example, if data variable $d_1$ is encoded using both length and a continuous colormap, and data variable $d_2$ is encoded using length only, we only need to consider $d_1$-length vs. $d_2$-length. We can define the five levels as:
\begin{enumerate}
    \item[5.] \emph{Major}: none; \emph{Medium}: none; \emph{Minor}: none.%
    \item[4.] \emph{Major}: none; \emph{Medium}: none; \emph{Minor}: one or a few.%
    \item[3.] \emph{Major}: none; \emph{Medium}: one; \emph{Minor}: more than a few.%
    \item[2.] \emph{Major}: none; \emph{Medium}: more than one; \emph{Minor}: any.%
    \item[1.] \emph{Major}: at least one. \emph{Medium}: any; \emph{Minor}: any.
\end{enumerate}
\noindent where we recommend that the term ``a few'' is defined as less than 10\% of all pairwise comparisons, and ``more than a few'' is 10\% or more but less than 50\%.
We also recommend to evaluate this criterion holistically by obtaining a Type D score directly. When there is no need for comparing any data variables within a glyph, we recommend to set the weight for this criterion to zero. See also Appendix \ref{apx:Comparability}.

\subsection{Attention: Importance (Criterion 8)}
\label{sec:Importance}
\begin{itemize}
\item \textbf{Definition:} This criterion assesses the desirable level of support in a glyph design for encoding data variables according to their importance, e.g., by allocating more pre-attentive visual channels or higher encoding bandwidth to more important data variables.%
\item \textbf{Recommended Modes:} \textcolor{gray}{Type B, Type C,} Type D (direct).%
\item \textbf{Recommended Weight:} Type D [0.5].
\end{itemize}

Ropinski and Preim \cite{ropinski2011survey} instigated the benefit of encoding data variables according to their importance in the context of an application. Maguire et al. \cite{Maguire:2012:TVCG} identified several factors that may influence importance ranking: the level of a variable in a taxonomy, its usage in users' tasks, and so on.
The factors that help a visual channel receive more attention include the pop-out effect, the hierarchy effect, the size of the visual objects hosting the visual channel, and so on. Furthermore, when a data variable is encoded using multiple visual channels, it will likely receive more attention.
Maguire et al. presented a method to bring the rankings of variables and visual channels together in glyph design.    

Chung et al. \cite{chung2015glyph} defined ``attention balance'' as a criterion for matching the levels of attention that visual channels may receive with the importance levels of the variables. The term ``attention balance'' implicitly indicates two sides of the same coin.
Following the third design consideration, we make these two sides as two sub-criteria. This criterion focuses on ``importance'', and the next criterion on ``balance''. When there is no importance ranking of the data variables within a glyph, we recommend to set the weight for the importance criterion to zero. Nevertheless, the balance criterion will always be assessed. 

\begin{figure}[t]
    \centering
    \includegraphics[width=\linewidth]{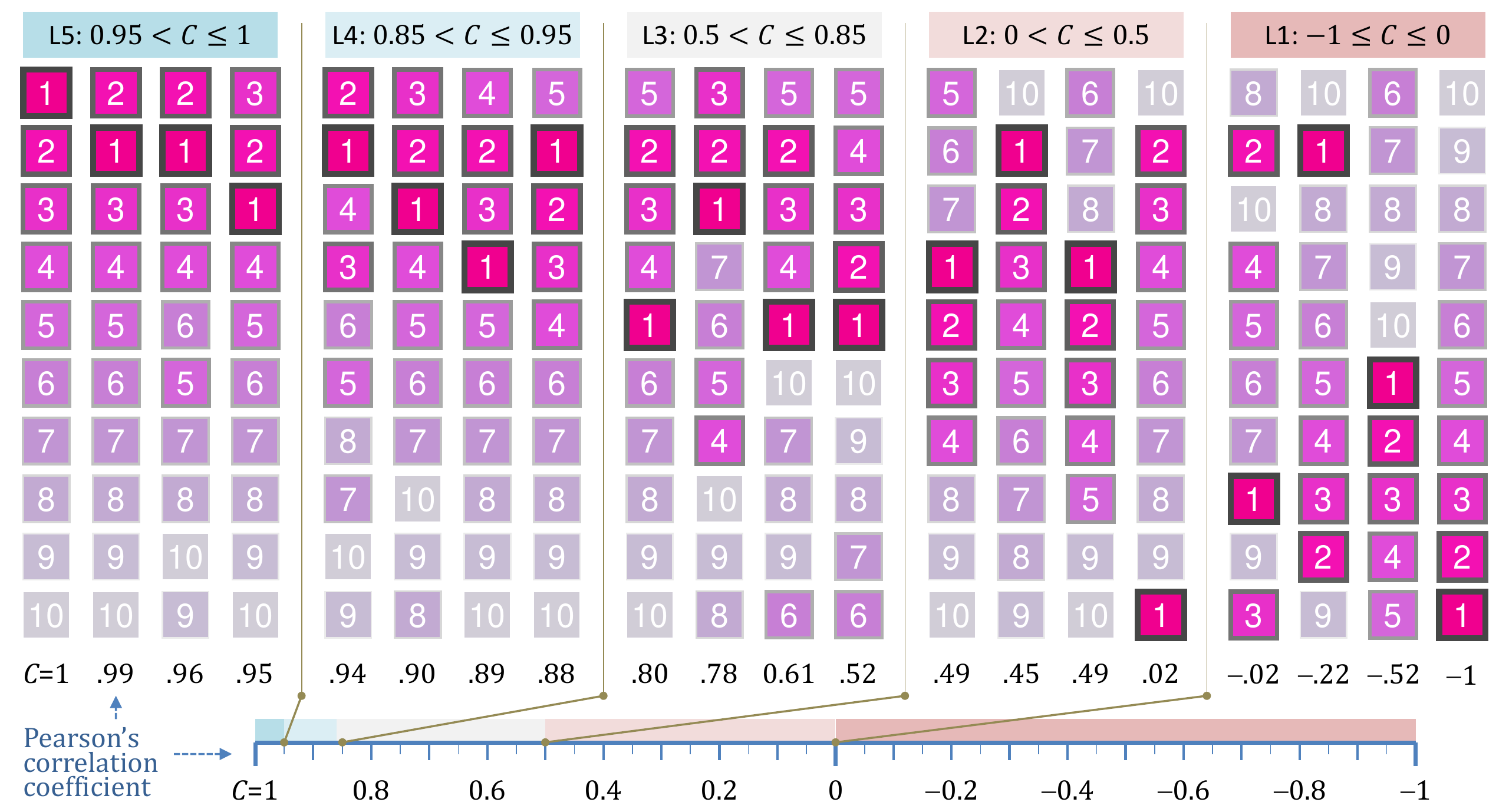}
    \caption{Five levels of the ``Attention: Importance'' criterion are defined based on different amounts of correlation between the importance ranks of data variables (encoded using $y$-position) and their attention ranks (encoded using number, color, edge thickness, and edge darkness). Four examples are shown at each level.}
    \label{fig:Importance}
\end{figure}

Consider a list of data variables, $d_1, d_2, \ldots, d_n$. Each variable $d_i$ is associated with two ranking values: $\iota_i$ for the importance of $d_i$ and $\alpha_i$ for the attention of $d_i$ through its visual encoding.
We have $\iota_i, \alpha_i \in [1, n]$.
The highest ranks of importance and attention are represented by 1 and the lowest by $n$. If two data variables are ranked same for importance (or attention), their $\iota$ (or $\alpha$) values are the same. We can compute the Pearson correlation coefficient (see also Appendix \ref{apx:Importance}):
%
\[
    C = \frac{\sum (\iota_i - \overline{\iota})(\alpha_i - \overline{\alpha}) }{\sqrt{\sum (\iota_i - \overline{\iota})^2 \sum (\alpha_i - \overline{\alpha})^2}}
\]
{\noindent where $\overline{\iota}$ and $\overline{\alpha}$ are mean ranking values of importance and attention respectively. The five levels are defined as:
\begin{enumerate}
    \item[5.] The correlation coefficient $C > 0.95$.%
    \item[4.] The correlation coefficient $0.85 < C \leq 0.95$.%
    \item[3.] The correlation coefficient $0.5 < C \leq 0.85$.%
    \item[2.] The correlation coefficient $0 < C \leq 0.5$.%
    \item[1.] The correlation coefficient $C \leq 0$.
\end{enumerate}

Fig.~\ref{fig:Importance} shows four examples at each level. We can observe minor misalignment of the ordering has limited impact on the correlation coefficient, which is suitable for the uncertainty in ranking importance and attention, because of the subjective nature of importance ranking and the lack of experimental measures in attention ranking.
In theory, one may first obtain Type B or Type C scores. In practice, it is more efficient to compute a Type D score directly.  

\begin{figure*}[ht]
    \centering
    \includegraphics[width=180mm]{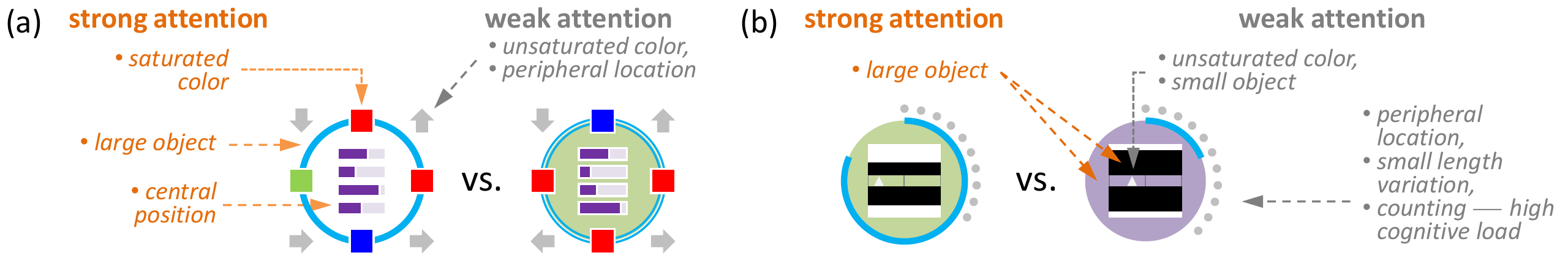}
    \caption{Examples of data variables that are ``overshadowed'' by other data variables, and may easily be overlooked (i.e., inattentional blindness). When one is asked to point out which variables have been changed between the two glyphs, the arrow directions in (a) and the triangular marker and the length of the dotted arc in (b) may receive significantly less attention than other variables. }
    \label{fig:Imbalance}
\end{figure*}
\subsection{Attention: Balance (Criterion 9)}
\label{sec:Balance}
\begin{itemize}
\item \textbf{Definition:} This criterion assesses the undesirable disadvantages that some data variables may suffer, which may make such data variables easily overlooked or difficult to perceive. 
\item \textbf{Recommended Modes:} \textcolor{gray}{Type B, Type C,} Type D (direct).%
\item \textbf{Recommended Weight:} Type D [0.5].%
\end{itemize}
As described in Section \ref{sec:Importance}, this criterion is assessing the opposite side of the same coin of ``Attention''. When attention is prioritized for the importance of data variables, it is necessary to ensure that no data variable may be seriously disadvantaged or suffer from inattentional blindness, which is a phenomenon studied extensively in psychology. In the context of glyph design, inattentional blindness primarily occurs when some data variables attract significantly more attention and thus limit cognitive resource, causing the variations of some other data variables to go unnoticed.

Here we assume that (i) the variables concerned are discernable (see Section \ref{sec:Discernability}), and (ii) the importance-based ordering is correct (see Section \ref{sec:Importance}). The blindness is caused by imbalanced allocation of cognitive resource for noticing variations. The factors of imbalance may include (a) peripheral location, (b) unsaturated color, (c) minor shape variation, (d) small object, (e) variation demanding high cognitive load, and so on. As illustrated in Fig.~\ref{fig:Imbalance}, the blindness is usually due to the co-existence of two or more such factors.

Ideally, empirical research in the future will provide us with methods for identifying visual encoding that may suffer from inattentional blindness. Until then, one may identify such a variable by juxtaposing two glyphs (of the same design) where all data variables have some variations. As illustrated in Fig.~\ref{fig:Imbalance}, one can observe those variables receiving weak attention, i.e., their variations are easily overshadowed by other variables.
We recommend evaluating this criterion holistically by obtaining a Type D score directly. We define the five levels based on the number of data variables that receive weak attention and may cause inattentional blindness:
\begin{enumerate}
    \item[5.] No data variable receives weak attention.%
    \item[4.] One data variable receives weak attention.%
    \item[3.] Two data variables receive weak attention.%
    \item[2.] Three data variables receive weak attention.%
    \item[1.] More than three data variables receive weak attention.
\end{enumerate}

\subsection{Searchability (Criterion 10)}
\label{sec:Searchability}
\begin{itemize}
\item \textbf{Definition:} This criterion assesses the desirable property that the visual channel(s) for each data variable can be recognized easily among others after a viewer has learned and remembered the encoding scheme. 
\item \textbf{Recommended Modes:} \textcolor{gray}{Type C,} Type D (direct).%
\item \textbf{Recommended Weight:} Type D [0.5].%
\end{itemize}
Chung et al. \cite{chung2015glyph} defined ``Searchability'' as the level of ease when one needs to identify a visual channel associated with a specific data variable. Here we assume that the user has already learned and remembered such an association semantically. This allows us to consider searchability independently of whether the encoding is easy to learn or remember. As illustrated in Fig.~\ref{fig:Searchability}, when many variables have similar visual encoding except their positions in a glyph, they can be difficult to find, despite their encoding following Bertin's rules and receiving adequate attention. See also Appendices \ref{apx:Orthogonality} and \ref{apx:Searchability}.

\begin{figure}[t]
    \centering
    \includegraphics[width=\linewidth]{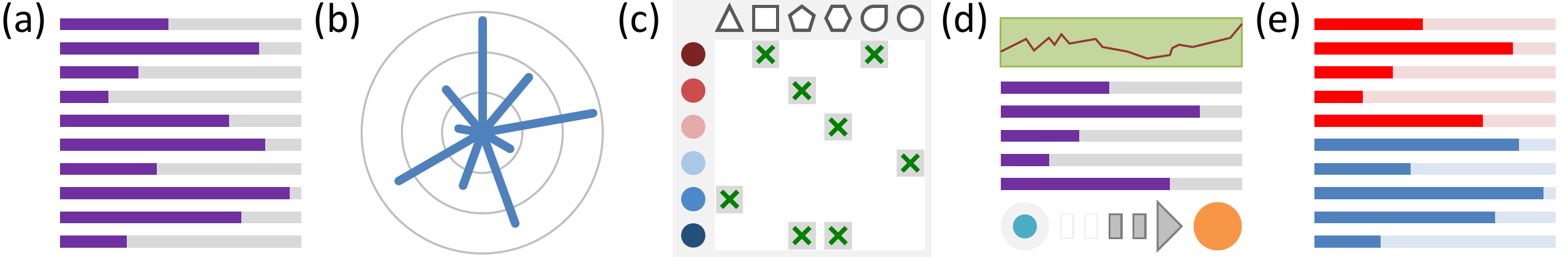}
    \caption{Three visual designs can encode (a) 10 numerical variables, (b) 9 numerical variables, and (c) 36 Boolean variables respectively. For some visual channels in these three glyphs, it is not easy to relate a visual channel to a specific variable, though they work fine in normal large plots. Problems can be alleviated if the number of similar visual channels are reduced as in (d) and (e).}
    \label{fig:Searchability}
\end{figure}

Let us consider a meta-variable $v_\text{meta} = r(d, \lambda)$ for representing the association between a data variable $d$ and a visual channel $\lambda$, such that $v_\text{meta} = \textbf{true}$ if an association exists and \textbf{false} otherwise. In each of the three examples in Fig.~\ref{fig:Searchability}, $v_\text{meta}$ is encoded primarily using a spacial location, which may also be searched through a related visual cue such as angle or count. In psychology, previous experiments have shown that the accuracy and response time of visual search and counting are affected by the number of objects and some other factors.
For example, in Fig.~\ref{fig:Searchability}(a), the first and last bars are easier to search than other eight. 
If there were a smaller number of bars, e.g., a group of five bars in Fig.~\ref{fig:Searchability}(d,e), symmetry can aid the visual search.
In Fig.~\ref{fig:Searchability}(b), for some numbers, the bars can be placed along lines from the center to the vertices of a square, hexagon, octagon, or even dodecagon, with one vertex at the 12 o'clock direction. These placements can also aid visual search. One can easily use multiple encoding by adding additional visual cues (e.g., colors and symbols) to improve the searchability. For example, in Fig.~\ref{fig:Searchability}(c), one could replace each cross symbol with an object that has the shape defining the column and the color defining the row. In Fig.~\ref{fig:Searchability}(e), the 10 bars are divided into two groups using two colors. 
Ideally, the assessment of this criterion could be based on the measurement of accuracy, response time, and/or cognitive load in visual search. The previous empirical research in psychology has not provided standardized measurements that can be used for assessing glyph designs. We anticipate that this will be obtained in future empirical studies in visualization. For the time being, we coarsely define three levels of cognitive effort in visual search as:

\begin{itemize}
    \item \emph{Low cognitive load} -- It requires almost no effort to find a specific variable, e.g., the first or last bar in Fig.~\ref{fig:Searchability}(a).%
    \item \emph{Medium cognitive load} -- It requires a small and undemanding amount of counting or reasoning effort. One normally feels such an effort, but is fairly sure about the search results. For example, the 2nd and 3rd bars in Fig.~\ref{fig:Searchability}(a) fall into this category.%
    \item \emph{High cognitive load} -- It requires an amount of searching effort that one feels bothersome or burdensome, while one may hesitate about the correctness of the search. For example, the 4th through the 7th bar in Fig.~\ref{fig:Searchability}(a) fall into this category.
\end{itemize}
Based on these three categories, we can define the five levels as:

\begin{enumerate}
    \item[5.] \emph{High}: none; \emph{Medium}: none; \emph{Low} all.%
    \item[4.] \emph{High}: none; \emph{Medium}: one or a few; \emph{Low}: most.%
    \item[3.] \emph{High}: none; \emph{Medium}: more than a few; \emph{Low}: more than half.%
    \item[2.] \emph{High}: one or a few; \emph{Medium}: any; \emph{Low}: any.%
    \item[1.] \emph{High}: more than a few; \emph{Medium}: any; \emph{Low}: any.
\end{enumerate}
\noindent where we recommend that the term “a few” is defined as fewer than 10\% of all data variables. We also recommend to evaluate this criterion holistically by obtaining a Type D score directly.

\subsection{Learnability (Criterion 11)}
\label{sec:Learnability}
\begin{itemize}
\item \textbf{Definition:} This criterion assesses the desirable property that the whole encoding scheme of a glyph is easy to explain and learn.%
\item \textbf{Recommended Modes:} Type D (direct).%
\item \textbf{Recommended Weight:} Type D [0.5].
\end{itemize}

Chung et al. \cite{chung2015glyph} defined learnability as the level of ease in learning and remembering a visual encoding scheme. As learning and memorizing are often studied separately in psychology, we split ``learnability'' and ``memorability'' into two related sub-criteria. For example, if the glyph in Fig.~\ref{fig:Searchability}(a) encodes the average marks of 10 courses (unnumbered), the scheme is easy to learn but difficult to remember. If it encodes the attendance of 10 weeks in an academic term, it is both easy to learn and remember.   
While both learnability and memorability can benefit from the intuitive encoding of individual visual channels (see Section \ref{sec:Intuitiveness}), there are many holistic factors, such as the total number of data variables, their relative positions, their semantic similarity and difference, and so on.
In order not to overload a criterion (Design Consideration 2), we let intuitiveness focus on individual visual channels through Type A scores, while focusing learnability and the memorability on two holistic sub-criteria through Type D scores.

Learnability is user-dependent. In order to focus on glyph designs, we assume that the target users have already had the knowledge about the data variables to be encoded, e.g.,
terms such as scrum, ruck, lineout, maul, and try in the context of rugby sports \cite{Legg:2012:CGF}.
In many applications, glyph-based visualization is designed for domain experts. Because of the assumption of domain knowledge, controlled or semi-controlled empirical studies are typically unsuitable for assessing this criterion since the lack of domain knowledge of the experiment participants would invalidate the experiment results.    
Meanwhile, learnability should be assessed in relation to the baseline that the target users have little knowledge about the visual design concerned. For example, domain experts often contribute directly ideas of visual encoding in a design process. Naturally, these domain experts have already ``learned'' the designs to be evaluated and their knowledge would bias the assessment.      

For a typical target user with adequate domain knowledge but little knowledge about the visual design to be evaluated, we define the five levels based on \emph{learning time}, \emph{learning mode} (i.e., levels of training engagement), and the effort required for \emph{repeated learning} after a short period of not using the glyphs. The five levels are:
\begin{enumerate}
    \item[5.] \emph{Learning time}: $<0.5$ hours; \emph{Learning mode}: self-learning only; \emph{Repeated learning}: effortless.%
    \item[4.] \emph{Learning time}: $\geq 0.5,\,< 1.0$ hour; \emph{Learning mode}: self-learning + Q\&A; \emph{Repeated learning}: effortless.%
    \item[3.] \emph{Learning time}: $\geq 1.0,\,< 1.5$ hours; \emph{Learning mode}: tutorial; \emph{Repeated learning}: minor effort.%
    \item[2.] \emph{Learning time}: $\geq 1.5,\,< 2.0$ hours; \emph{Learning mode}: tutorial; \emph{Repeated learning}: noticeable effort.%
    \item[1.] \emph{Learning time}: $\geq 3$ hours; \emph{Learning mode}: tutorial; \emph{Repeated learning}: serious effort.%
\end{enumerate}
\noindent Here we define three learning modes: self-learning only, self-learning + Q\&A, and tutorial. We define the effort for repeated learning as effortless (e.g., a quick glance at the encoding scheme), minor effort (e.g., reading the encoding scheme again for 5--10 minutes), noticeable (e.g., reading the encoding scheme again for 10--30 minutes and/or requiring Q\&A), and serious effort (e.g., requiring another tutorial and/or more than 30 minutes).  
Note that the frequency of repeated learning relates to memorability. See also Appendix \ref{apx:Learnability}.

\subsection{Memorability (Criterion 12)}
\label{sec:Memorability}
\begin{itemize}
\item \textbf{Definition:} This criterion assesses the desirable property that the whole encoding scheme of a glyph is easy to remember once a viewer has learned the scheme.%
\item \textbf{Recommended Modes:} Type D (direct).%
\item \textbf{Recommended Weight:} Type D [0.5].
\end{itemize}
As already discussed in Section \ref{sec:Learnability}, this complementary criterion assesses the easiness of memorizing an encoding scheme. Similar to learnability, it is a holistic criterion and is assessed through a Type D score. The assessment assumes that the users have already learned the encoding scheme, and the effort for repeated learning and memory refreshing is considered as part of learnability.

\begin{table*}[t]
\centering
\caption{A summary of the assessments of five glyph designs using the MCDA-aided scheme. \textbf{A} is is the original design by Maguire et al. \cite{Maguire:2012:TVCG}.
\textbf{B} is a variant of \textbf{A} configured based on several design options discussed in \cite{Maguire:2012:TVCG}.   
\textbf{C} is the original design by Legg et al. \cite{Legg:2012:CGF}.
\textbf{D} is a variant of \textbf{C} where two visual objects swap their positions and the central pictograms are replaced with abstract shapes discussed in \cite{Legg:2012:CGF}. 
\textbf{E} is a design by Chung et al. \cite{chung2015glyph}, which was partly based on \textbf{C}.
Only Type D scores are shown. Further details, including Type A scores, can be found in a spreadsheet file in the supplementary material.}
\label{tab:CaseStudy1}
%
\begin{tabular}{@{\hspace{48mm}}c@{\hspace{6mm}}c@{\hspace{12mm}}c@{\hspace{8mm}}c@{\hspace{11mm}}c@{}}
    \includegraphics[height=10mm]{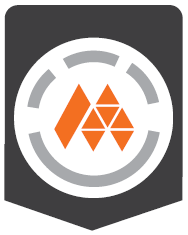} &
    \includegraphics[height=10mm]{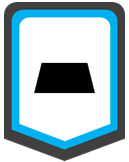} &
    \includegraphics[height=10mm]{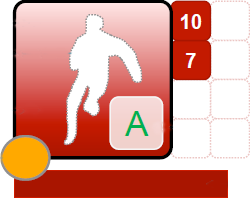} &
    \includegraphics[height=10mm]{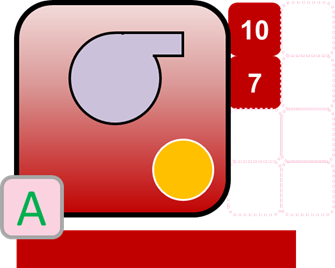} &
    \includegraphics[height=10mm]{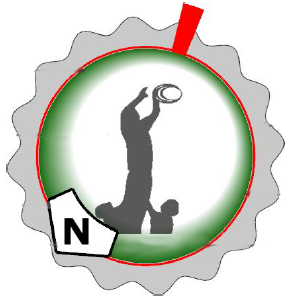} \\
    \textbf{A:} Maguire et al. &
    \textbf{B:} Parody of \textbf{A} &
    \textbf{C:} Legg et al. &
    \textbf{D:} Parody of \textbf{C} &
    \textbf{E:} Chung et al.\\
\end{tabular}
\begin{tabular}{@{}p{46mm}@{\hspace{4mm}}%
        c@{\hspace{3mm}}c@{\hspace{8mm}}c@{\hspace{3mm}}c@{\hspace{12mm}}%
        c@{\hspace{3mm}}c@{\hspace{8mm}}c@{\hspace{3mm}}c@{\hspace{12mm}}c@{\hspace{3mm}}c@{}}
    \textbf{Criterion} &
        \textbf{weight} & \textbf{score} & \textbf{weight} & \textbf{score} &
        \textbf{weight} & \textbf{score} & \textbf{weight} & \textbf{score} &
        \textbf{weight} & \textbf{score}\\
    \hline
    Typedness                  &   1 & 5.00 &   1 & 4.71 &   1 & 5.00 &    1 & 5.00 &   1 & 5.00\\
    Discernability             &   1 & 5.00 &   1 & 5.00 &   1 & 5.00 &    1 & 5.00 &   1 & 5.00\\
    Intuitiveness              &   1 & 4.14 &   1 & 3.29 &   1 & 4.13 &    1 & 3.63 &   1 & 4.10\\
    Invariance: Geometry       & 0.5 &    5 & 0.5 &    4 & 0.5 &    5 &  0.5 &    5 & 0.5 &    3\\
    Invariance: Colorimetry    & 0.5 &    3 & 0.5 &    3 & 0.5 &    5 &  0.5 &    5 & 0.5 &    4\\
    Composition: Separability  & 0.5 &    5 & 0.5 &    1 & 0.5 &    5 &  0.5 &    3 & 0.5 &    5\\
    Composition: Comparability &     &      &     &      &     &      &      &      &     &\\
    Attention: Importance      & 0.5 &    5 & 0.5 &    5 & 0.5 &    5 &  0.5 &    4 & 0.5 &    5\\
    Attention: Balance         & 0.5 &    5 & 0.5 &    2 & 0.5 &    5 &  0.5 &    5 & 0.5 &    5\\
    Searchability              & 0.5 &    5 & 0.5 &    1 & 0.5 &    5 &  0.5 &    5 & 0.5 &    5\\
    Learnability               & 0.5 &    5 & 0.5 &    2 & 0.5 &    5 &  0.5 &    3 & 0.5 &    4\\
    Memorability               & 0.5 &    4 & 0.5 &    1 & 0.5 &    5 &  0.5 &    1 & 0.5 &    3\\
    \hline
    \textbf{Total Weight \& Weighted Average}
                               &   7 & 4.66 & 7   & 3.21 & 7   & 4.80 & 7    & 4.16 & 7    & 4.44\\ 
\end{tabular}
\end{table*}

Because of the effort to learn a glyph design, the learned encoding scheme must be stored in long-term memory. While one could assess how long the target users can remember an encoding scheme, the overall intention of this MCDA method is to evaluate different glyph designs without too much delay. Therefore, we recommend to base the assessment of this criterion on the memorability after 1 hour and 24 hours following learning. Both time periods meet the requirement for testing long-term memory \cite{Glanzer:1966:JVLVB,Baddeley:2020:book}. See also Appendix \ref{apx:Memorability}.

For a typical target user, we define the five levels according to how much the user can remember about an encoding scheme:

\begin{enumerate}
    \item[5.] \emph{after 1 hour}: 100\%, and \emph{after 24 hours}: 100\%.%
    \item[4.] \emph{after 1 hour}: $<100\%$, $\geq 90\%$, or\\ \emph{after 24 hours}: $<100\%$, $\geq 75\%$.%
    \item[3.] \emph{after 1 hour}: $<90\%$, $\geq 75\%$, or\\ \emph{after 24 hours}: $<75\%$, $\geq 50\%$.%
    \item[2.] \emph{after 1 hour}: $<75\%$, $\geq 50\%$, or\\ \emph{after 24 hours}: $<50\%$, $\geq 25\%$.%
    \item[1.] \emph{after 1 hour}: $<50\%$, or\\ \emph{after 24 hours}: $<25\%$.
\end{enumerate}

\begin{table*}[t]
\centering
\caption{A summary of the assessments of five optional glyph designs using the MCDA-aided scheme.
$J_1$ uses six lines to represent the min-max ranges of the six angles.
$J_2$ combines each pair of lines for the two related angles into a single line. $J_3$ is similar to $J_2$ but uses 2D arcs instead of lines.
$J_4$ is similar to $J_3$ but uses 3D arcs instead of 2D arcs. 
$J_5$ is organizes the six arcs around a circle, and indicates three pairs using colored wedges. \reviseTVCG{The scores shown are the mean values of two assessors' scores.}}
\label{tab:CaseStudy2}
%
\begin{tabular}{@{\hspace{23mm}}c@{\hspace{8mm}}c@{\hspace{15mm}}c@{\hspace{13mm}}c@{\hspace{12mm}}c@{\hspace{12mm}}c@{}}
    &
    \includegraphics[height=10mm]{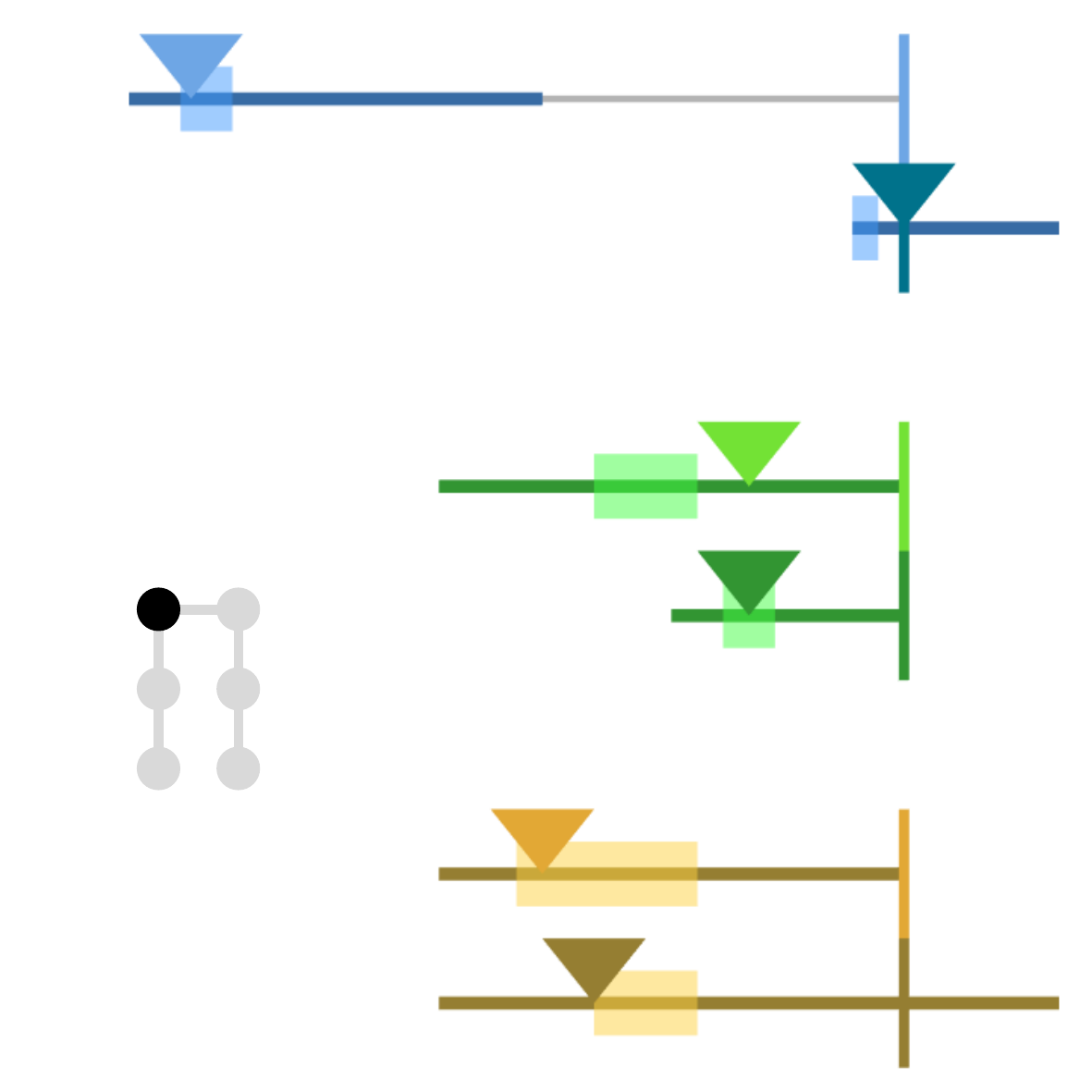} &
    \includegraphics[height=10mm]{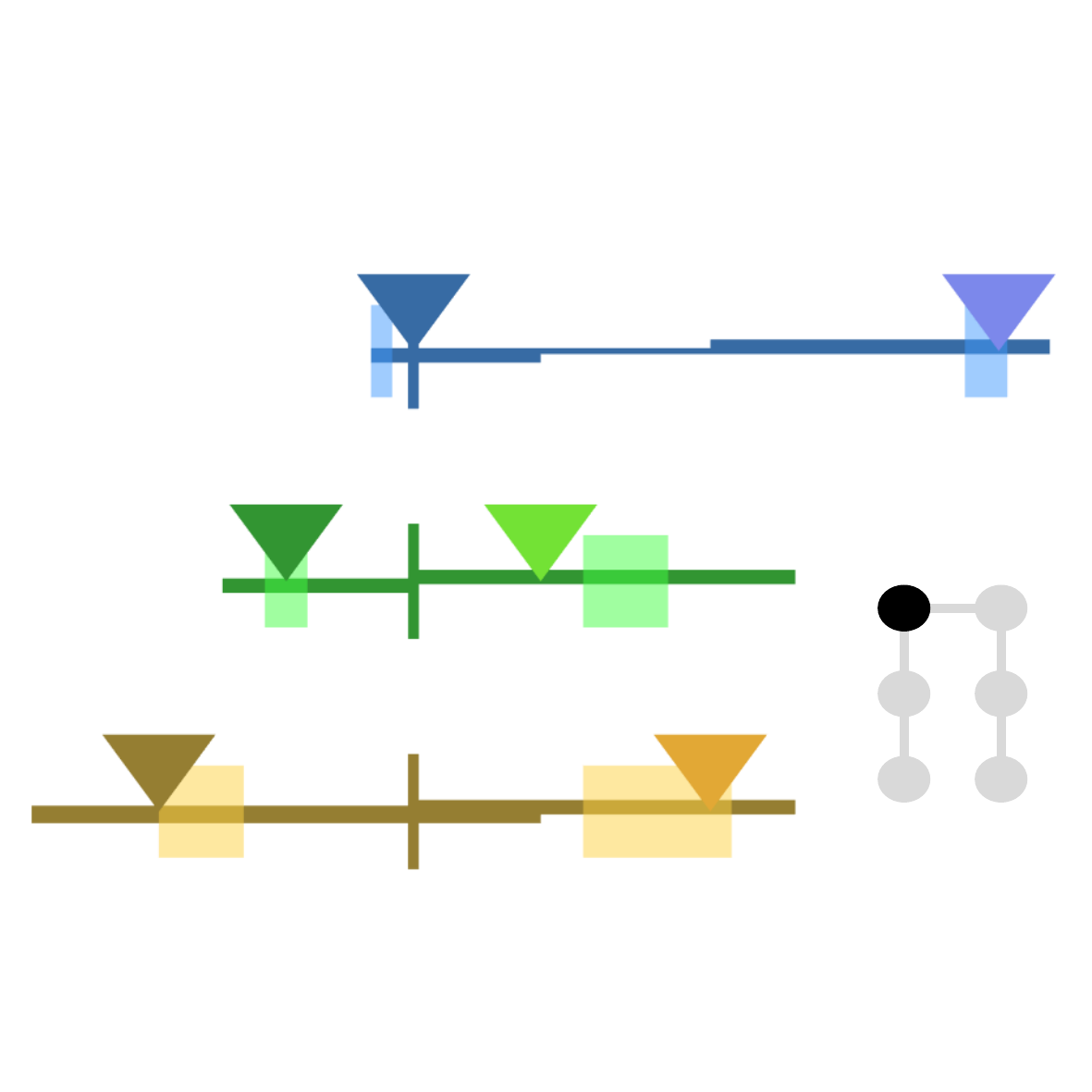} &
    \includegraphics[height=10mm]{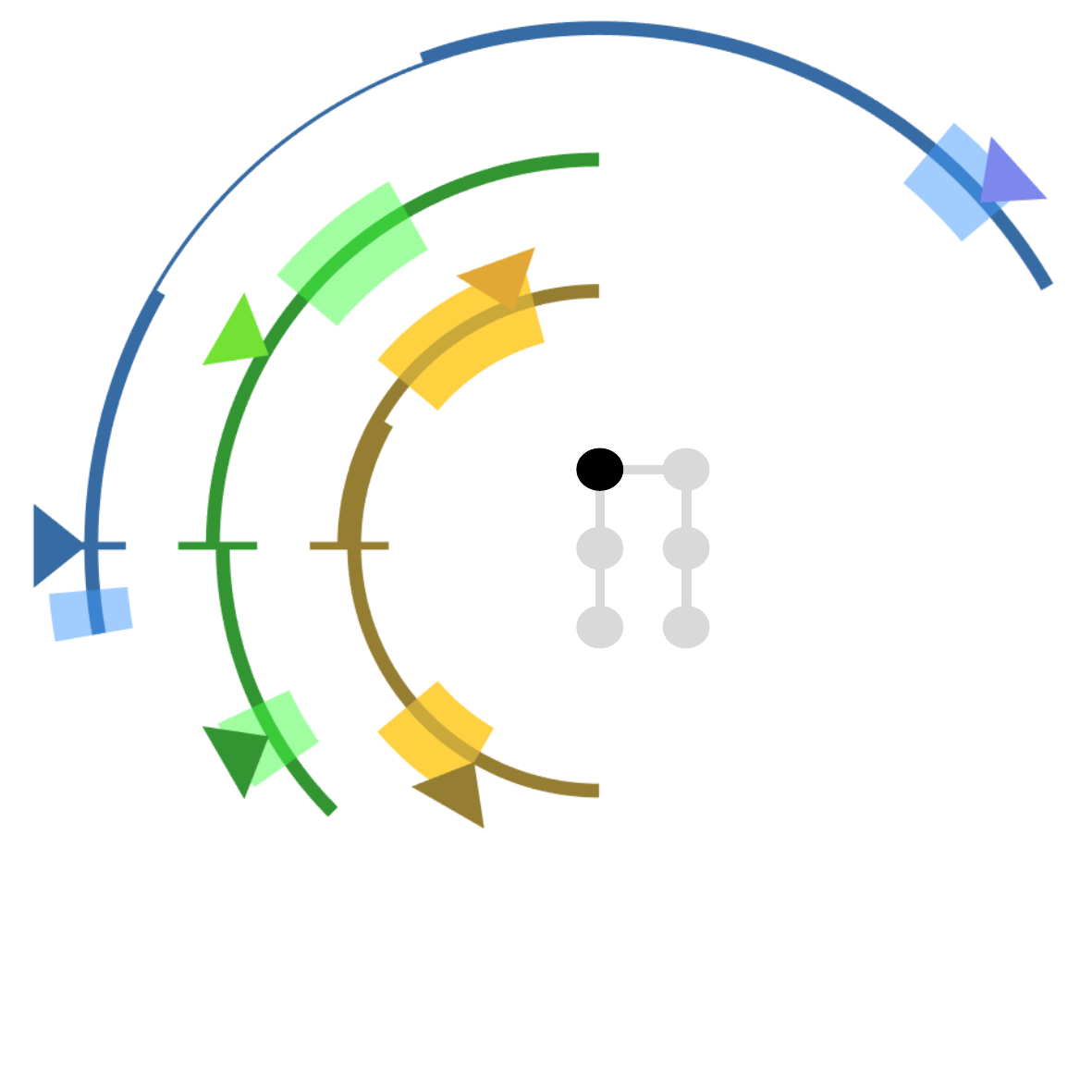} &
    \includegraphics[height=10mm]{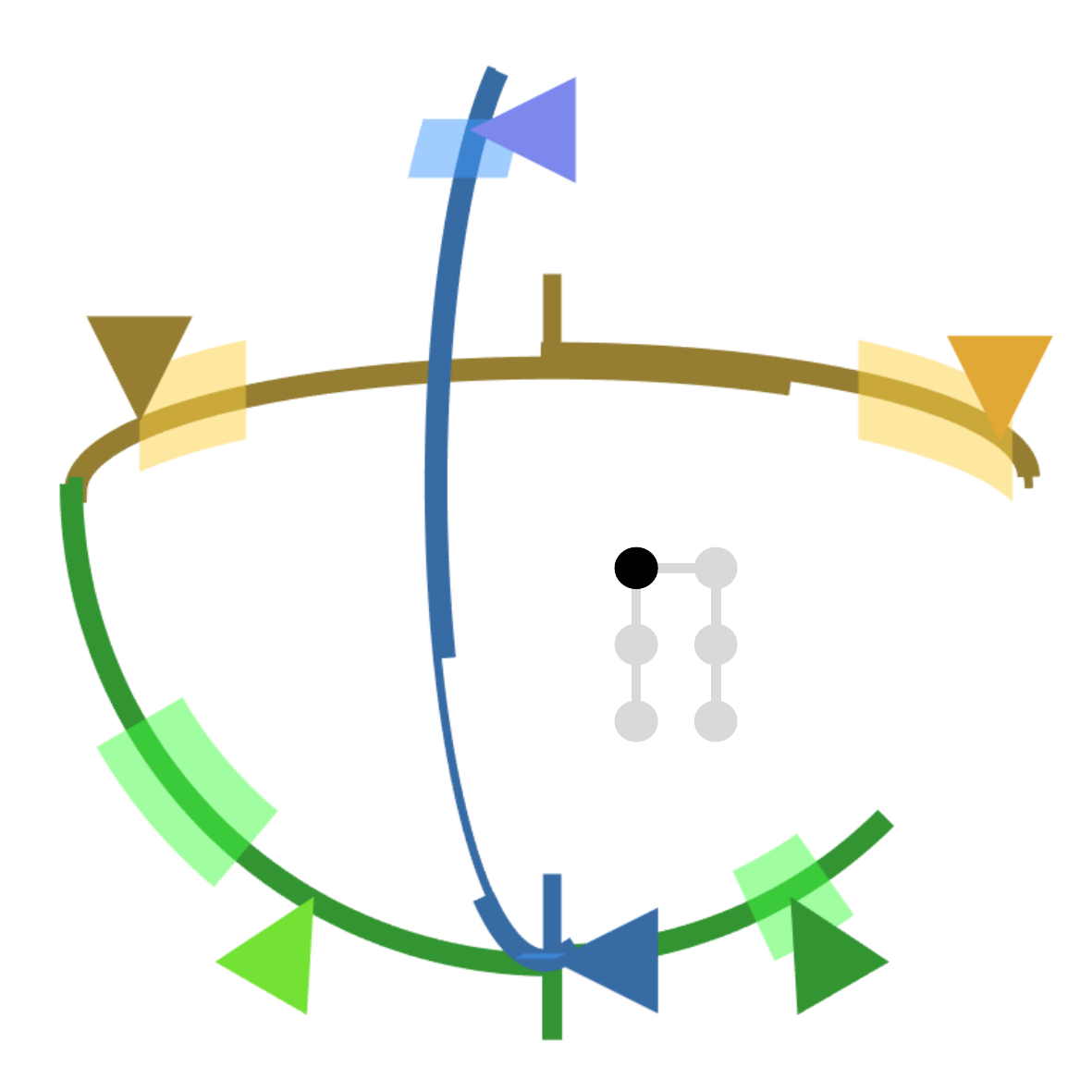} &
    \includegraphics[height=10mm]{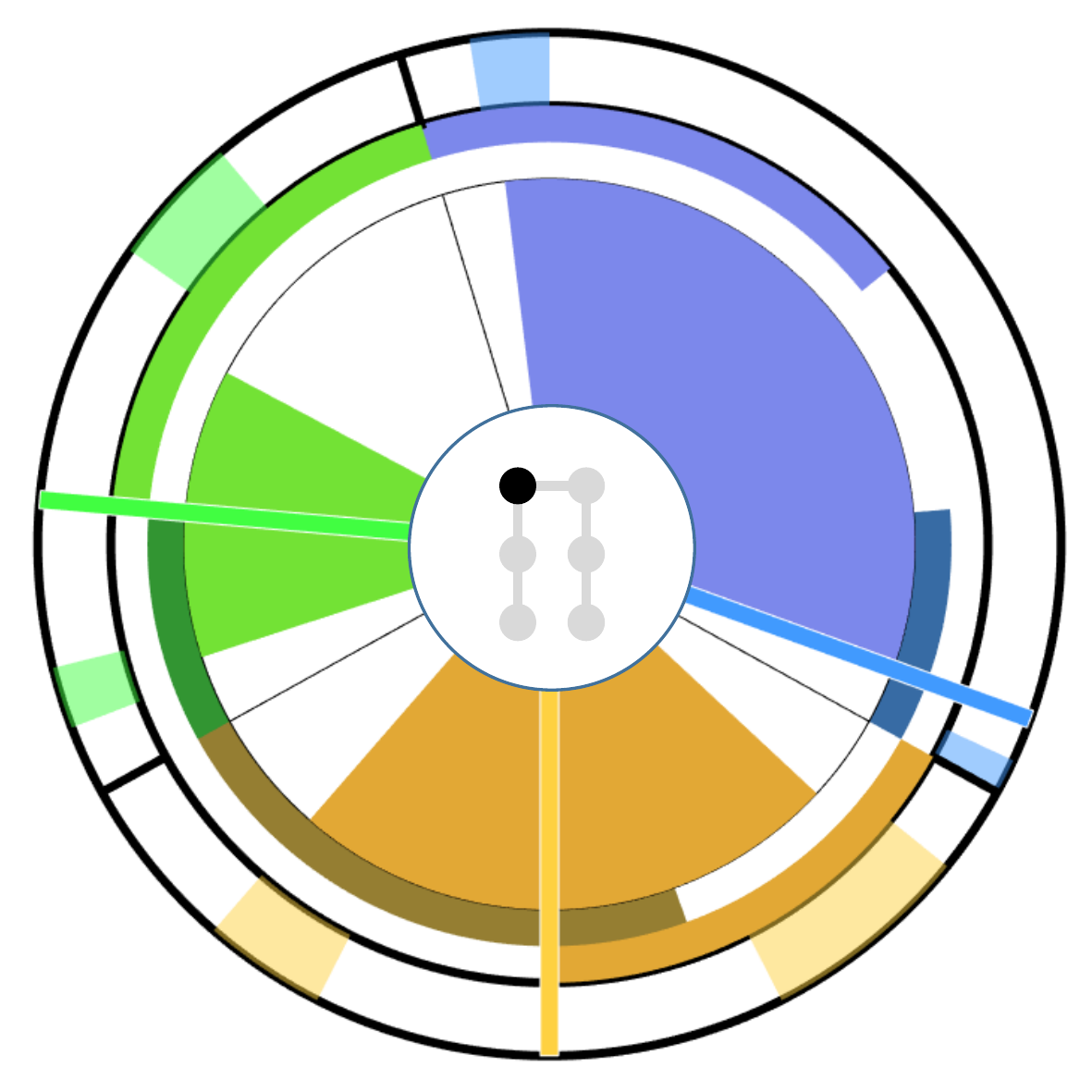} \\
    \textbf{Design Options:} &
    \textbf{$J_1$: 6-line} &
    \textbf{$J_2$: 3-line} &
    \textbf{$J_3$ 3-arc 2D}&
    \textbf{$J_4$: 3-arc 3D} &
    \textbf{$J_5$: Pie-arc} \\
\end{tabular}
\begin{tabular}{@{}p{46mm}@{\hspace{4mm}}%
        c@{\hspace{3mm}}c@{\hspace{9mm}}c@{\hspace{3mm}}c@{\hspace{9mm}}%
        c@{\hspace{3mm}}c@{\hspace{9mm}}c@{\hspace{3mm}}c@{\hspace{9mm}}c@{\hspace{3mm}}c@{}}
    \textbf{Criterion} &
        \textbf{weight} & \textbf{score} & \textbf{weight} & \textbf{score} &
        \textbf{weight} & \textbf{score} & \textbf{weight} & \textbf{score} &
        \textbf{weight} & \textbf{score}\\
    \hline
    Typedness                  &   1 & 4.84 &   1 & 4.84 &   1 & 4.84 &   1 & 4.84 &   1 & 4.92\\
    Discernability             &   1 & 4.60 &   1 & 3.14 &   1 & 3.46 &   1 & 3.46 &   1 & 3.68\\
    Intuitiveness              &   1 & 3.43 &   1 & 3.03 &   1 & 4.81 &   1 & 4.97 &   1 & 2.27\\
    Invariance: Geometry       & 0.5 &  4.5 & 0.5 &  4.5 & 0.5 &    5 & 0.5 &    4 & 0.5 &    5\\
    Invariance: Colorimetry    & 0.5 &    5 & 0.5 &    5 & 0.5 &    5 & 0.5 &    5 & 0.5 &    5\\
    Composition: Separability  & 0.5 &    5 & 0.5 &    3 & 0.5 &    3 & 0.5 &    3 & 0.5 &    3\\
    Composition: Comparability & 0.5 &    5 & 0.5 &    4 & 0.5 &    5 & 0.5 &    3 & 0.5 &    3\\
    Attention: Importance      & 0.5 &    5 & 0.5 &    5 & 0.5 &    5 & 0.5 &    5 & 0.5 &    5\\
    Attention: Balance         & 0.5 &    5 & 0.5 &    5 & 0.5 &    5 & 0.5 &    5 & 0.5 &    5\\
    Searchability              & 0.5 &  4.5 & 0.5 &  4.5 & 0.5 &  4.5 & 0.5 &    5 & 0.5 &    5\\
    Learnability               & 0.5 &  3.5 & 0.5 &  3.5 & 0.5 &    4 & 0.5 &    5 & 0.5 &    2\\
    Memorability               & 0.5 &    4 & 0.5 &  3.5 & 0.5 &    4 & 0.5 &    5 & 0.5 &    3\\
    \hline
    \textbf{Total Weight \& Weighted Average}
                               & 7.5 & 4.48 & 7.5 & 4.00 & 7.5 & 4.45 & 7.5 & 4.44 & 7.5 & 3.85\\ 
\end{tabular}
\end{table*}

\section{MCDA-aided Evaluation: \reviseTVCG{Scoring Examples}}
\label{sec:Evaluation}
We applied the MCDA scheme described in Section \ref{sec:Scheme} to a number of visual designs in the literature and their ``parodies'' (alternative designs) and to some new designs for a biomechanical application. These \reviseTVCG{scoring examples} allowed us to test the scheme, identify ambiguous definitions, unbalanced categorization, and inappropriate thresholds, facilitating the improvement of the scheme. We have included three files (two design documents and one spreadsheet workbook) in the supplementary materials.

\vspace{1mm}
\noindent\textbf{Existing Designs and Parodies}.
Table \ref{tab:CaseStudy1} shows the summary of five \reviseTVCG{examples}. 
Among these, \textbf{A} is the original design by Maguire et al. \cite{Maguire:2012:TVCG}, who also reported a number of alternative encoding methods for individual data variables. It is not difficult for us to configure ``parody'' designs based on these alternative encoding methods. Design \textbf{B} is one of such parody designs. In \textbf{B}, variable S6 is encoded using three colors instead of three shapes (e.g., dark gray instead white circle in Table \ref{tab:CaseStudy1}). Variable S2 is encoded using five colors for an outline instead of five metaphoric shapes (e.g., cyan square instead of an icon for material combination). S5 is encoded using seven basic shapes instead of seven countable metaphoric shapes as shown in the first row of Fig.~\ref{fig:Intuitiveness}.     
From Table \ref{tab:CaseStudy1}, we can observe that the parody design has over-used colors and basic shapes, having a negative impact on several criteria, especially in terms of separability, attention balance, searchability, learnability, and memorability.

\textbf{C} is the original design by Legg et al. for supporting real-time event analysis during a rugby match. Design \textbf{D} is a parody design of \textbf{C} with two modifications: (i) replacing the silhouette pictogram at the center of the glyph with abstract shapes, and (ii) swapping the locations of the outcome circle (orange for unsuccessful) with the territory box (location A) in the glyphs representing Design \textbf{C} and Design \textbf{D}. As the data variable for event types has 16 key values, in comparison with pictograms in \textbf{C}, the abstract shapes in \textbf{D} reduces intuitiveness, learnability, and memorability. Meanwhile, the swapping changes the order between variables outcome and territory. The outcome circle is not as attentive as in \textbf{C} and becomes less separable from the abstract shapes in the center.

\textbf{E} is a design by Chung et al. \cite{chung2015glyph}, where a similar set of pictograms was used for event types. Unlike \textbf{C} that was designed for rugby coaches and sports analysts, \textbf{E} was designed for analysts only, and it contains several numerical variables resulting from video analysis, such as gain, tortuosity, and net lateral movement. The discernability of these variables are affected by size and color degeneration. Although visual designs for these variables were introduced for this application, the analysts who had the technical background could learn and memorize these with a bit of extra effort.

We have included the documentation about the parody designs \textbf{B} and \textbf{D} in the supplementary materials.
\reviseTVCG{There were many scoring exercises that authors have conducted throughout this work. This exercise was among several exercises conducted after the MCDA scheme in Section \ref{sec:Scheme} was more or less finalized. These exercises were designed to debug the text of various definitions in the scheme, and to synchronize the interpretation among the authors. In this particular exercise, two authors first scored the five designs independently. Three authors then met and discussed the two sets of scores. They reached an agreement on every score. The final scores are presented in Table \ref{tab:CaseStudy1}. The detailed Type A scores are provided in the supplementary material. We recommend that ``discussion meeting'' is used as the default approach when involving multiple assessors, especially when some assessors are less familiar with the MCDA scheme.}


\vspace{1mm}
\noindent\textbf{New Designs for a Biomechanical Application}. This \reviseTVCG{example} is a design exercise for studying the feasibility of using glyph-based visualization in a medical application. This design exercise considered a subset of data variables, i.e., those related to rotations.    
In musculoskeletal biomechanics and related fields, such as physiotherapy, sports science, and orthopaedics, it is common to measure the angular motion of the large joints of the body (e.g shoulder or knee). The rotations can be quantified in three dimensions using Euler angles: typically called \emph{flexion-extension}, \emph{abduction-adduction}, and \emph{internal-external} rotation. Each of these angles has both maximum and minimum possible values and a ``normal'' range in each rotation direction, giving 24 \emph{populational variables} in total for each joint. When an individual is examined, one or more measures are usually obtained for each angle, giving at least 6 \emph{individual variables} for each joint. When the values of these variables are recorded for an individual in a data table, there are additional labels indicating a few categorical values, such as joint type (neck, hip, knee, ankle, etc.), body side (left, right, and center), and rotation axes and directions. To the best of our knowledge, the total number of variables is much higher than those in the existing glyph designs in the literature (e.g., $\sim$20 variables in Duffy et al. \cite{Duffy:2015:TVCG}).

\revise{While data tables are customarily used in musculoskeletal biomechanics and related fields, using glyphs to encode these data tables is helpful when domain experts need to observe and analyze a number of data tables, e.g., monitoring treatment progress, comparing different clusters, and identifying commonality and anomalies in a cohort.}

The \reviseTVCG{example} is part of a project, which involves an expert of biomechanics, an expert of visualization, and a postgraduate student working on the intersection of the two subjects. During a period of a few years, numerous glyph designs were created for different sets of biomechanical variables. This particular \reviseTVCG{example} focuses on the rotational measures. Several dozens of designs were created for encoding the aforementioned individual and populational variables. These designs were evaluated qualitatively in several meetings, and five designs were selected to be examined in detail using the MCDA scheme described in Section \ref{sec:Scheme}. Table \ref{tab:CaseStudy2} shows the summary of five designs in this \reviseTVCG{example}. The scores for ``Typedness'', ``Discernability'', and ``Intuitiveness'' are Type D scores, each was aggregated from 37 Type A scores.

The scores were first produced by the two VIS researchers independently, and then discussed in several project meetings, where the biomechanics expert offered advisory comments on the scores. \reviseTVCG{The authors noticed that their initial scores were largely in agreement and considered that using mean values would be a valid approach.} A few original scores were adjusted independently by the assessors concerned, and the mean values were calculated as presented in Table \ref{tab:CaseStudy2}. \reviseTVCG{For most criteria, they arrived the same score for each design. There was no difference of more than 1 level. In other words, every integer score resulted from two independent scores that were the same. Each 4.5 score resulted from two close scores 4 and 5, while each 3.5 score resulted from two close scores 3 and 4. The two sets of independent scores (including the detailed Type A scores) are provided in the supplementary material.}

\reviseTVCG{This scoring exercise was conducted many months after the exercise reported in Table \ref{tab:CaseStudy1}. By then, the wordings of the definitions in Section \ref{sec:Scheme} had gone through several iterations of refinement and the authors' interpretation of the text had become fairly consistent. ``Averaging'' is thus the second approach when involving multiple experienced assessors.}

\begin{figure}
    \centering
    \begin{tabular}{@{}c@{\hspace{2mm}}c@{\hspace{2mm}}c@{}}
         \includegraphics[width=26mm]{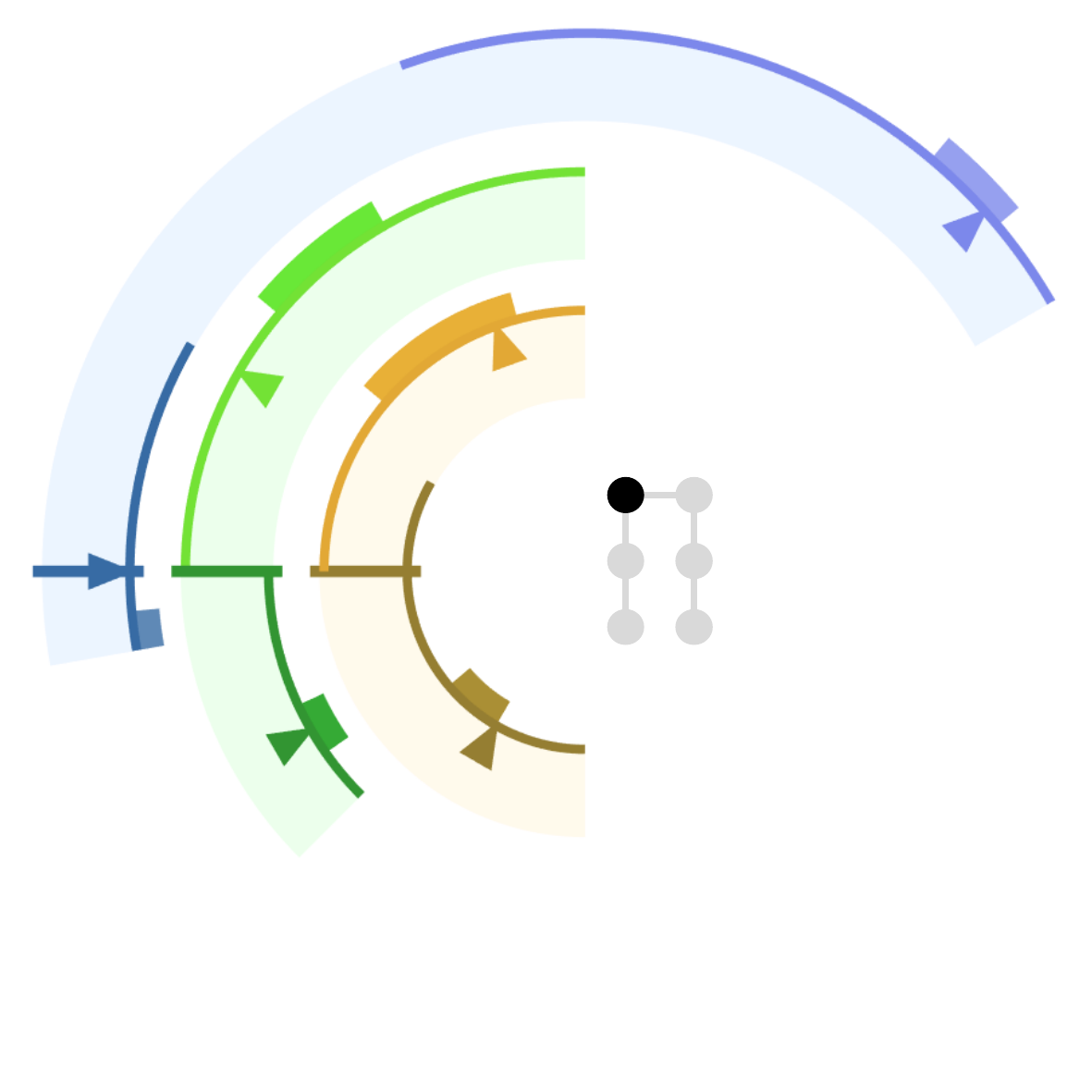} &
         \includegraphics[width=26mm]{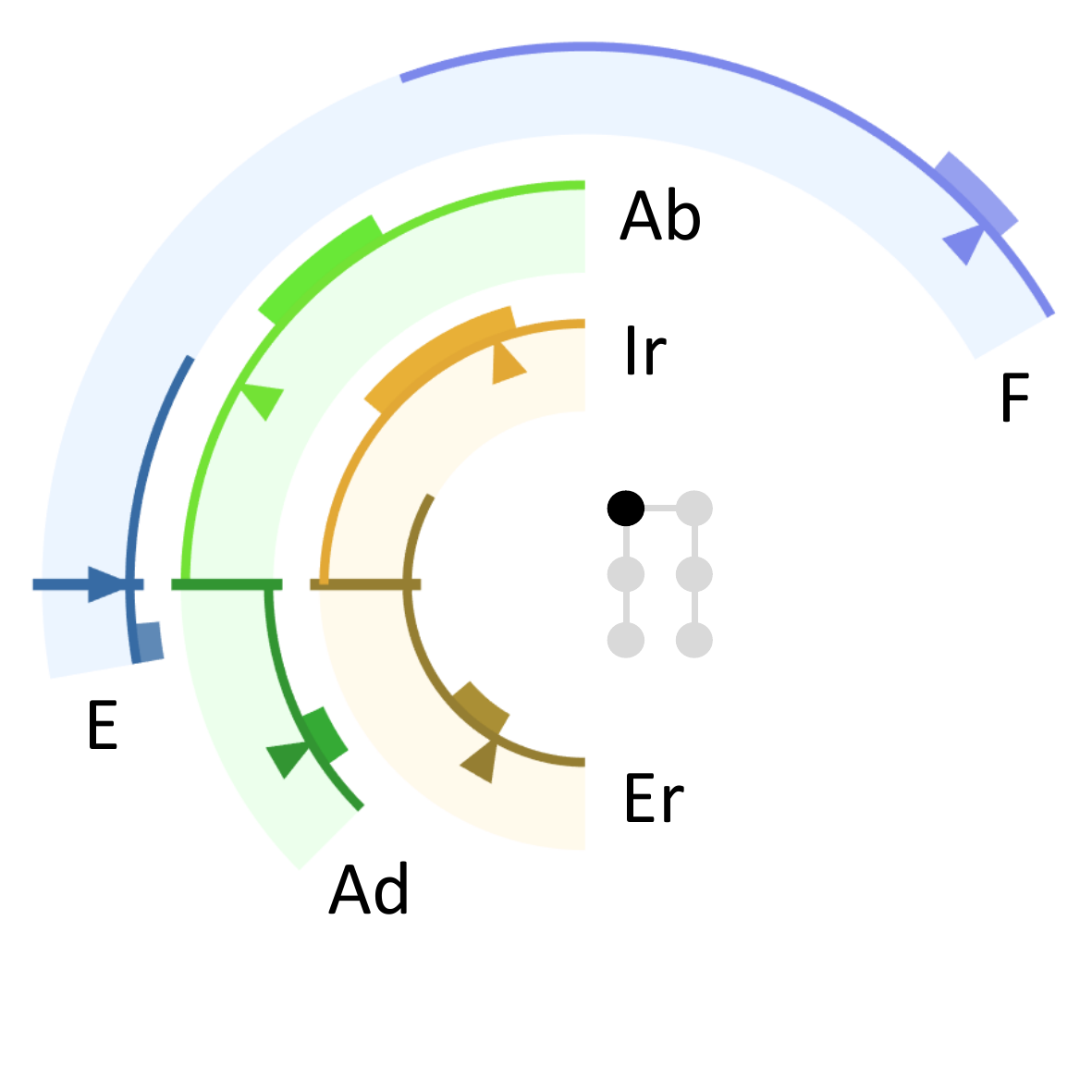} &
         \includegraphics[width=26mm]{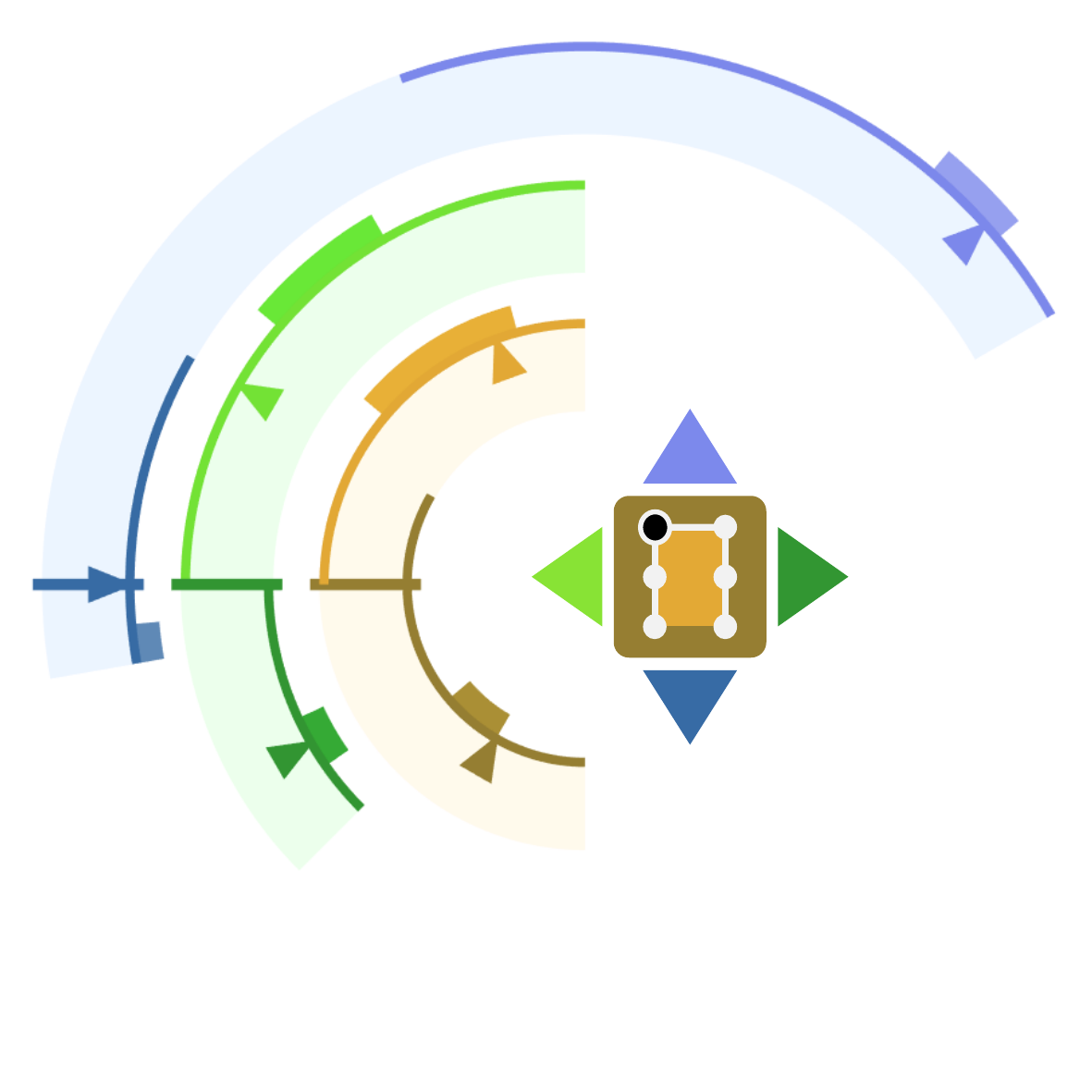} \\[-4mm]
         \small{(a) $J_6$ 6-arc 2D Design} &
         \small{(b) $J_{6}$ with labels} &
         \small{(c) $J_{6}$ with legend}
    \end{tabular}
    \caption{Two further designs that were created after the assessment of the five designs in Table \ref{tab:CaseStudy2}. (a) Design $J_6$ was based on $J_1$ and $J_3$ and received an overall score 4.73. (b) Each arc is associated an abbreviated label for the corresponding rotation angle. (c) A mini-legend shows six color-coded directions. The overall scores of (b) and (c) are about 4.80.}
    \label{fig:FurtherDesigns}
\end{figure}

The overall scores ordered the five designs as $J_1 (4.48)$, $J_3 (4.45)$, $J_4 (4.44)$, $J_2 (4.00)$, $J_5 (3.85)$. All team members agreed with the ordering.
While the scores of MCDA assessment should not be treated as a ground truth, the scores can help identify the shortcomings in the existing designs, including the best-ranked design.
\reviseTVCG{For example, although Designs $J_1$, $J_3$, and $J_4$ received similar high overall scores, each has its own strengthens and weaknesses. $J_1$ is strong on its ``Discernability'' but weak on ``Intuitiveness''. $J_4$ is noticeably strong on ``Intuitiveness'', ``Learnability'', and ``Memorability'' because the 3D layout depicts the physical meanings of the three rotation angles, but is noticeably weak on ``Separability'' and ``Comparability'' also because of the 3D layout.}

The better ``Discernability'' of $J_1$ directed us to a common shortcoming of the 3-line and 3-arc designs ($J_2, J_3$), where the two related angles may have overlapped ranges for some joints. When overlapping occurs, the depiction of some values can be confusing. We therefore introduced a 6-arc 2D design $J_6$ as shown in Fig. \ref{fig:FurtherDesigns}(a). It was based on $J_3$ by incorporating the 6-line idea in $J_1$ and bringing together the merits of $J_1$ in terms of ``Discernability'' and ``Separability'' and that of $J_3$ in terms of ``Intuitiveness''. $J_6$ received an overall score 4.73, indicating an improvement over $J_1$ (4.48) and $J_3$ (4.45).  

\reviseTVCG{The strength of Design $J_4$ in ``Intuitiveness'',  ``Learnability'' and ``Memorability'' directed us to consider labeling as an alternative to 3D layout.}
We considered several ways for labeling the six arcs, including using text labels and a mini-legend as shown in Figs. \ref{fig:FurtherDesigns}(b,c) respectively. Our MCDA scores indicated that both designs in Figs. \ref{fig:FurtherDesigns}(b,c) received approximately the same overall score $\approx 4.80$. Our analysis showed that text labels are easier to learn and demand less memorization than a mini-legend, resulting in a slightly higher score in terms of ``Learnability'' and ``Memorability'' than the latter. However, labels are affected by geometric scaling more than the arcs, arrows, and normal bands, resulting in a lower score in terms of ``invariance: Geometry''. In different biomechanical applications, users have different levels of expertise about the motion of joints and may use glyph-based visualization at different levels of frequency. We decide to keep all three design options in Fig. \ref{fig:FurtherDesigns}, and to finalize the designs for individual applications by assessing the requirements for ``Learnability'' and ``Memorability'' based on users' expertise and frequency of user tasks.

\section{Discussions and Conclusions}
\label{sec:Conclusions}
\textbf{Summary.} In this work, we have formulated a MCDA-aided assessment scheme for supporting glyph designs, and have tested the scheme on a range of glyph designs in the literature as well as their variants configured in the testing process. The scheme is built on the existing qualitative criteria in the literature (see also Appendices \ref{apx:FrameworkDesign} and \ref{apx:Bertin}), and enables a major step forward towards a more systematic, consistent, and semi-quantitative approach. Even when a glyph designer does not rank each criterion quantitatively, the scheme can serve as a reminder of the major considerations in encoding data variables and integrating different visual channels into a glyph representation. We also outline detailed workflows for assessing these criteria (Appendix \ref{apx:Workflow}).

\vspace{1mm}
\noindent \textbf{Limitations.}
(a) The scheme does not encode any domain-specific information. It is not in any way a replacement for user-centered design and evaluation, especially when the target users are domain experts. Nevertheless, the scheme can help glyph designers phrase questions in seeking advice from domain experts and can speed up the design process.
(b) The weights recommended in this work are specified based on our tests and analysis. On the one hand, there is a need for a set of weights that can be consistently applied to most (if not all) applications. On the other hand, weights are not ground truth, and their optimization needs the participation of the VIS community through many iterations. Thus the current recommendation is expected to be improved in the future. Meanwhile, it is important for glyph designers to ensure that the weights used in the assessment are transparent.
(c) For several criteria, e.g., separability, comparability, attention balance, and searchability, the specification of the five levels can be improved in the future based on new empirical research designed to obtain more precise measurements. We hope that the proposed scheme will stimulate such research.
(d) The proposed scheme does not cover the designs of 3D glyphs, glyph layout, interaction with glyphs, multi-scale glyphs, and so on. We hope that future research will extend the scheme.  

\vspace{1mm}
\noindent \textbf{Future Work.} In addition to the aforementioned future research areas, the field of VIS would benefit from publicly-available glyph editing tools. The proposed scheme could potentially be integrated into such tools, facilitating human-centered and semi-automatic assessment of glyph designs during the design process.

\bibliographystyle{abbrv-doi-hyperref}
\bibliography{main}

\clearpage


\renewcommand\thesection{\arabic{section}}
\renewcommand\thesubsection{\thesection.\Alph{subsection}}
\setcounter{section}{0}

\begin{center}
\large{APPENDICES}\\[2mm]
\Large\textbf{Multi-Criteria Decision Analysis\\
  for Aiding Glyph Design}\\[2mm]
\large Hong-Po Hsieh, Amy Zavatsky, and Min Chen\\
\end{center} 
\normalsize

\section*{}
The paper, which is entitled ``Multi-Criteria Decision Analysis for Aiding Glyph Design'', is accompanied by three appendices (in this document) and three further documents in the supplementary materials:
\begin{itemize}
    \item Appendix \ref{apx:FrameworkDesign} --- Providing an analysis of different categorization schemes in the literature and discussing why the proposed categorization scheme is suitable for the methodology of multi-criteria decision analysis. \reviseTVCG{It also includes discussions on the strategies for defining the five levels of each criterion.} 
    \item Appendix \ref{apx:Bertin} --- Providing the background information about the term ``Typedness'' and Bertin's criteria.
    \item Appendix \ref{apx:Workflow} --- Providing 12 detailed workflows, one for each criterion defined in Section \ref{sec:Scheme}.
    \item PDF File (ParodiesJointsDesigns.pdf) --- Providing detailed documentation about the parody designs and the glyph designs for a biomechanical application, which were reported in Section \ref{sec:Evaluation}.
    \item Spreadsheet File (GlyphRating\_ScoreSheets.xlsx) --- Providing multiple score sheets, each of which was used to score one of the designs in Section \ref{sec:Evaluation}. The spreadsheet file can also be used as a template for future applications of the MCDA scheme.  
\end{itemize}

\section{The Process for Formulating the Framework}
\label{apx:FrameworkDesign}

\begin{figure*}[h]
    \centering
    \includegraphics[width=\linewidth]{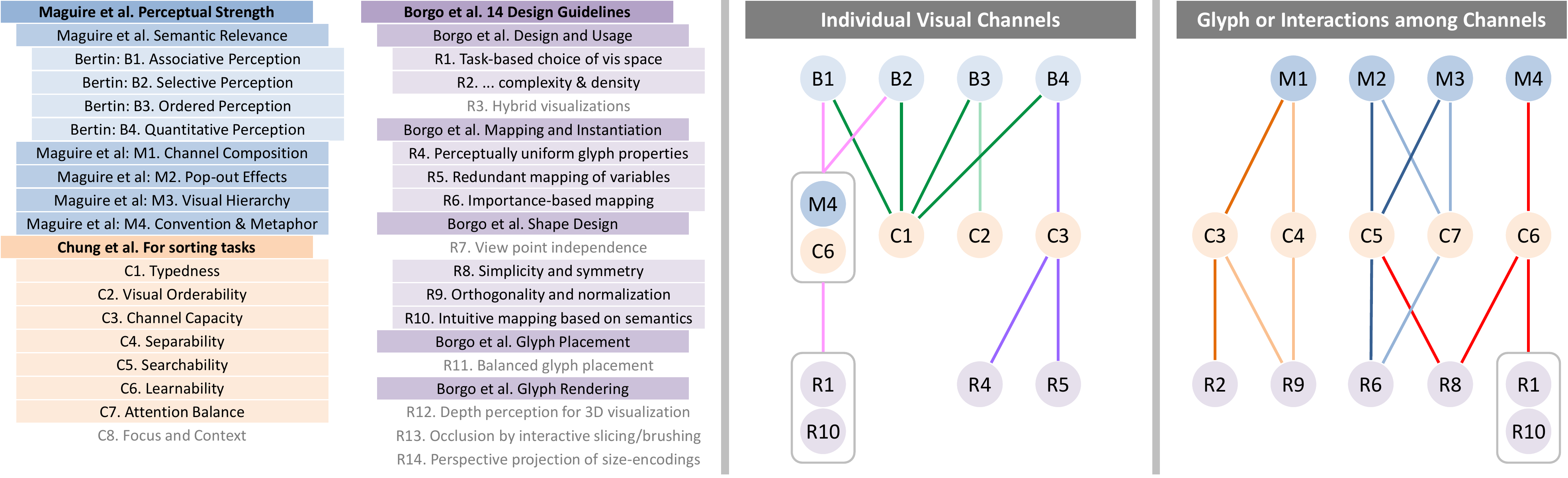}
    \caption{Left: The criteria proposed by Bertin \cite{Bertin:2011:book}, Maguire et al. \cite{Maguire:2012:TVCG}, Borgo et al. \cite{Borgo:2013:STAR}, and Chung et al. \cite{chung2015glyph}. Right: the connections among these criteria. Each connection line indicates partial overlapping between the two connected criteria.} 
    \label{fig:Criteria}
\end{figure*}

The members of the author team have different amounts of experience in glyph-based visualization, varying from working on glyph designs for about two years to co-authorship of some 20 papers involving glyph-related work, including three papers on the criteria for glyph designs (i.e., \cite{Maguire:2012:TVCG,Borgo:2013:STAR,chung2015glyph}, which are considered extensively in this work). During the processes of designing glyph-based visualization for a medical application, the team recognized that the design criteria in the literature (e.g., \cite{Bertin:2011:book,Maguire:2012:TVCG,Borgo:2013:STAR,chung2015glyph}) are not easy to use, because:
\begin{enumerate}
    \item Different sets of criteria are overlapping in some aspects but distinctive in other aspects. Fig. \ref{fig:Criteria} illustrates the overlapping and distinctive aspects. From the figure, we can also have a sense that combining these four sets is a not a trivial undertaking.
    \item Some criteria are defined with phrases ``should ideally'', ``may minimize'', and ``as much as possible'', reflecting the challenges for glyph designs to meet all criteria.
    \item On the one hand, one may wish to have a small set of criteria. However, each criterion would include several sub-criteria, making the combined criterion difficult to assess. One may also wish to have a set of highly-independent (often referred to as orthogonal) criteria, avoiding the same aspect being assessed multiple time. In general, narrowly-defined criteria are more likely to be independent of one another. However, the more narrowly-defined criteria are, the higher the number of criteria will be.  
    %
\end{enumerate}

The author team recognized the need to identify a set of criteria, each of which can be assessed with multi-level scores that transform semantic definitions of quality or desirability to numerical values. The operations research method ``multi-criteria decision analysis (MCDA)'' became the focus of this work, which commenced in October 2021. 

From Fig. \ref{fig:Criteria}, we can imagine many different options for combining these criteria, not mentioning further options which would be introduced in a combinatory fashion if one added new criteria or split some existing criteria. While Bertin's four levels of organization ($\equiv, \neq, \text{O}, \text{Q}$) \cite{Bertin:2011:book} may be translated into levels of quality or desirability, they cannot be applied to most (if not all) other criteria in Fig. \ref{fig:Criteria}, which were not accompanied by any level-definitions. When each criterion is enriched by defining such levels, there will be many different options for defining such levels. When an MCDA scheme is proposed for a set of criteria, there will be many options for assigning weighting values. The design space for such a framework is thus huge. Because the space is not a mathematically-defined space, one cannot utilize techniques such as mathematical optimization or machine learning to explore such a space.

Therefore, like almost all MCDA schemes in the literature, it will be a community effort to improve earlier schemes gradually. For this reason, we consider the criteria proposed by Bertin, Maguire et al., Borgo et al., and Chung et al. positively and highly valuable, rather than negatively as schemes that might not be optimized, since optimization is simply too challenging to obtain or ascertain.         


Since January 2022, the author team (including one member who devoted almost 100\% of research time on proposing this MCDA scheme) conducted a number of research activities, including:
\begin{itemize}
    \item Studying the literature on glyph-based visualization, aspects of perception and cognition, and MDCA methodology and applications. This was summarized in Section \ref{sec:RelatedWork}.
    \item Analysing the merits and potential shortcomings of each existing set of criteria, i.e., the Bs, Ms, Cs, and Rs in Fig. \ref{fig:Criteria} in the context of a premise that we were to transform the set to an MDCA scheme. We will describe these merits and shortcomings briefly in the four subsections below.
    \item Proposing a number of possible schemes and comparing their relative merits through both conceptual analysis and practice-based evaluation (i.e., by using each proposed scheme to score some glyph designs in the literature and their parody designs as well as our own designs for a set of biomechanical measures). We identified the best scheme among various other schemes considered including those based on the four existing sets of criteria. Because of the page limit, the conceptual analysis was concisely reported for each criterion individually in Section \ref{sec:Scheme}. The practice-based evaluation was selectively reported in Section \ref{sec:Evaluation}.
\end{itemize}

In Sections \ref{sec:BertinSet}$\sim$\ref{sec:BorgoSet}, we report our analysis of the merits and potential shortcomings of the four existing sets of criteria in the context of a premise that we were to transform the set to an MDCA scheme. This is not to be confused with an analysis of their relative merits in a broader context or in other contexts. In general, we do not make orthogonality among criteria an essential requirement since this not easily achievable. The difficulty is likely due to the fact that many different criteria share the same underlying perceptual and cognitive factors, or their underlying factors are intertwined. Therefore, we will point out such overlapping not as a criticism but as a potential angle for reducing overlapping through more precise definitions or reorganization of the grouping of criteria.

The discussions on the existing sets of criteria are followed by Section \ref{sec:BuildOn} where we summarize the major attributes in the existing sets that were adopted in our MCDA scheme described in Section \ref{sec:Scheme}.  

\subsection{The Four Criteria by Bertin}
\label{sec:BertinSet}
Bertin's four criteria are much better known in comparison to the other three sets of criteria. They are also accompanied by four levels of organization, i.e., selective ($\equiv$), associative ($\neq$), ordered (O) and quantitative (Q). \reviseTVCG{These can be translated to four different criteria in an MCDA scheme. When considering a visual channel against one of the four criteria, Bertin used binary assessment, i.e., yes or no (see also Appendix \ref{apx:Bertin})}. 

One may also recognize that there is some correlation between \emph{B1. Associative Perception} and \emph{B2. Selective Perception}, because of which, many have found it hard to distinguish between these two criteria. Meanwhile \emph{B3. Ordered Perception} and \emph{B4. Quantitative Perception} have some overlapping since a visual channel that is quantitative is also ordered.
However, Bertin made a serious effort to provide precise definitions of each criteria, which alleviated the confusion between overlapping criteria. 

On the other hand, as shown in Fig. \ref{fig:Criteria}, Bertin's four criteria may give a reasonable coverage of different aspects of an individual visual channel, but they do not cover some desirable qualities in terms of the interaction between different visual channels, such as \emph{C4. Searchability} and \emph{C7 Attention Balance} \cite{chung2015glyph}, and \emph{R8 Simplicity and Symmetry} and \emph{R9. Orthogonality and Normalization} \cite{Borgo:2013:STAR}.

\subsection{The Four Extra Criteria by Maguire et al.}
\label{sec:MaguireSet}
Maguire et al. \cite{Maguire:2012:TVCG} recognized the specific focus of \emph{B1}$\sim$\emph{B4} on visual channels individually, and proposed four additional criteria for evaluating desirable qualities in terms of the interaction among different visual channels. Among these, assessing \emph{M1. Channel Composition}, \emph{M2. Pop-out Effects}, and \emph{M3. Visual Hierarchy} involves the consideration of at least two visual channels, while \emph{M4. Convention \& Metaphor} may be applied to a visual channel individually or a whole glyph holistically.

Maguire et al. proposed \emph{M1}, \emph{M2}, and \emph{M3} according to some experimental findings and theoretical discourse in psychology, while proposing \emph{M4} based on their own experience of graphic design. Later Borgo et al. underpinned \emph{M4} with evidence from VIS literature and related it to memorability of a glyph design.  

On the other hand, the criteria proposed by Maguire et al. do not adequately cover the desirable quality for each visual channel to encode a sufficient number of data values, which is referred to as \emph{C3. Channel Capacity} by Chung et al. \cite{chung2015glyph}. Despite this lack, \emph{M1} captures an aspect of \emph{C3} as integration of visual channels may reduce their channel capacities. While \emph{M1} considers potential negative impact because of integrated composition, Maguire et al. did not consider the need for an effective composition in some glyph designs to facilitate comparative observation of different visual channels.     
Although Maguire et al. used redundant mapping in their glyph design, they did not propose a criterion to convey \emph{R5. Redundant mapping of variables} as done by Bogor et al. \cite{Borgo:2013:STAR}. Maguire et al. also conducted scalability tests, but did not include scalability as a criterion.    

\subsection{The Eight Criteria by Chung et al.}
\label{sec:ChungSet}
The reason that we discuss the eight criteria by Chung et al. \cite{chung2015glyph} before the fourteen criteria by Borgo et al.  \cite{Borgo:2013:STAR} is because the latter discussed the work by the former, though Chung et al. \cite{chung2015glyph} was published later.

Chung et al. combined Bertin's four criteria into a single criterion \emph{C1.
Typedness}, and introduced \emph{C3. Channel Capacity} as a new criterion.
Chung et al. used the terms ``$\square$-ability'' for four of their eight criteria, suggesting a possible intention to assess such criteria using a rating scale, such as a Likert-type scale.
For example, \emph{C4. Separability} is a rateable criterion corresponding to Maguire et al.'s \emph{M1}. \emph{C6. Learnability} is a rateable criterion corresponding to Maguire et al.'s \emph{M4}. Chung et al. recognized the pros and cons of pop-out effects (\emph{M2}) and visual hierarchy (\emph{M3}) and replaced them with \emph{C5. Searchability} and \emph{C7. Attention Balance}.

The criterion \emph{C2. Visual Orderability}, which is strongly related to their research on the topic \cite{Chung:2016:CGF}, overlaps with Bertin's \emph{B3} and \emph{B4}. The definition of \emph{C6. Learnability} includes both learning and remembering, but memorability is not obvious in the term learnability. Chung et al. did not include either scalability or comparability as a criterion.

For this work, we did not include the consideration of \emph{C8. Focus and Context} as this was for assessing glyphs to be used in interactive visualization. 

\subsection{The Fourteen Criteria by Borgo et al.}
\label{sec:BorgoSet}
Borgo et al. \cite{Borgo:2013:STAR} conducted a comprehensive survey on a broad range of topics in glyph-based visualization, including the theoretic foundations, design guidelines, techniques, and applications. They summarized a set of fourteen criteria as design guidelines. Here we focus on eight of them as the other six are for 3D glyph design, using glyphs in other visual representations (e.g., volume rendering), and glyph placement and occlusion when multiple glyphs are used in a visualization (e.g., on a geographical map). 

As shown in Fig. \ref{fig:Criteria}, the eight criteria proposed by Borgo et al., i.e., \emph{R1}, \emph{R2}, \emph{R4}, \emph{R5}, \emph{R6}, \emph{R8}, \emph{R9}, and \emph{R10}, are somehow related to the criteria proposed by Maguire et al. \cite{Maguire:2012:TVCG} and Chung et al. \cite{chung2015glyph}. For example, \emph{R4. Perceptually-uniform glyph properties} and \emph{R5. Redundant mapping of variables} are two approaches for improving \emph{C3. Channel Capacity}. \emph{R10. Intuitive mapping based on semantics} and \emph{R1. Task-based choice of visualization space} are related directly to \emph{C6} while broadening \emph{M4}. Criterion \emph{R6. Importance-based mapping} is related to \emph{C5} and \emph{C7} as well as \emph{M2} and \emph{M3}.

 Borgo et al. introduced the criterion \emph{R5. Redundant mapping of variables}, acknowledging the potential positive impact of using multiple visual channels to encode a data variable.

Unlike Maguire et al. and Chung et al., Borgo et al. did not include Bertin's four criteria in their categorization, nor were the criteria scalability and comparability considered. Because they defined their categorization as a set of design guidelines, some guidelines are stated in a way ``to try to do if one can'', which is not quite the same as criteria to be assessed. 

\subsection{Building on the Existing Sets of Criteria}
\label{sec:BuildOn}
The development of the four sets of criteria by Bertin, Maguire et al., Chung et al., and Borgo et al. reflects a research-based process for improving our scientific understanding about glyph-based visualization as well as practical methods for evaluating glyph designs. While we anticipate that this research-based process will continue for years and decades, it is important for us to continue this process, building on the existing contributions, and making further improvement. Here we summarize the main attributes that we can adopt or adapt in formulating an MCDA scheme.
\begin{itemize}
    \item \textbf{Rateable Criteria} --- Bertin's criteria are rateable (see Appendix \ref{apx:Bertin}). Chung et al. used terms ``$\square$-ability'', which suggest the potential for defining rateable criteria.
    \item \textbf{Channel-focused vs. Glyph-focused} --- From Fig. \ref{fig:Criteria}, we can observe that some criteria focus on the desired qualities of individual visual channels, and some focus on the whole glyph or the interaction among visual channels. This difference entails assessment at different levels of detail. Some criteria can be used to assess both channel-focused and glyph-focused qualities.
    \item \textbf{Major Clusters of Criteria} --- From Fig. \ref{fig:Criteria}, we can observe the major clusters of criteria connected by lines colored differently. These include: (i) Bertin's criteria (two shades of green), (ii) channel capacity (purple), (iii) channel intuitiveness (pink), (iv) composition and separability (two shades of orange), (v) attention and searchability (two shades of blue), and (vi) glyph intuitiveness (red).
    \item \textbf{Missing Criteria} --- The missing criteria identified in the above sections, i.e., scalability and comparability, should be included in the new scheme.
    \item \textbf{Two Sides of the Same Coin} --- Chung et al. considered the positive and negative impact of varying attention in terms of \emph{C5} and \emph{C7}. One may consider the two sides of varying composition in terms of interference (\emph{M1}) and comparability.
    \item \textbf{Sub-criteria} --- Some differences among different sets of criteria in Fig. \ref{fig:Criteria} reflect the desire to assess sub-criteria separately. There is a trade-off between assessing a criterion holistically and assessing its sub-criteria individually.
    \item \textbf{One-to-many Mapping} --- Borgo et al. introduced the potential use of multiple visual channels to encode an important data variable as a guideline, while Maguire et al. demonstrated the merits of such one-to-many encoding. In fact, one-to-many encoding is more common than one expects and often cannot be avoided. For example, a circular arc naturally has at least several visual channels: arc length, arc angle, starting and terminating angles in relation to 0 degrees. When a data value is encoded using an arc, one can perceive this value using both arc length and arc angle. Varying the height of a triangle causes changes to several visual channels including height, area, shape, and angles. Hence this raises a question: shall we assess each visual channel individually or assess all visual channels used to encode a data variable? 
\end{itemize}

Building on these attributes, the author team proposed a number of MCDA schemes. After some lengthy comparative analysis, these different MCDA schemes evolved to become the MCDA scheme described in Section \ref{sec:Scheme}.

\reviseTVCG{As briefly discussed at the beginning of Appendix \ref{apx:FrameworkDesign}, when an MCDA scheme is proposed for a set of criteria, there will be many options for assigning
weighting values. The design space for such a framework is thus huge. Because the space is not a mathematically-defined space, one cannot utilize techniques such as mathematical optimization to explore such a space. After the scheme have been used in tens and hundreds of design processes, the designers in the VIS community may share their experience and some colleagues may bring such experience together into coherent evaluation documents, which may lead to further improvement of the scheme, different weight distribution, or proposals of new schemes.}

\reviseTVCG{The current scheme is accompanied by a simple weight distribution. Criteria 1, 2, and 3 are Type A criteria (assessed through examining data variables individually). All of them are assigned a weight of 1 at the Type D level. Criteria 4 $\sim$ 12 are all Type D criteria and are assigned a weight of 0.5 at the Type D level. We can also observe that many criteria are in pairs (e.g., criteria [4, 5], [6, 7], [8, 9], and [11, 12], and this provides another justification for a 0.5 weight for each of these criteria. Meanwhile, we consider, qualitatively, that the impact level of criterion 10 is similar to those of criteria $4 \sim 9, 11 \sim 12$, while Criteria 1, 2, and 3 may have more critical impact than these criteria. This provides another justification for the difference between the 1.0 and 0.5 weights.}

\reviseTVCG{At the moment, there is no definite reason to make small adjustment of the weights (e.g., 0.4 instead of 0.5) for any criterion. We also do not recommend for designers to change the weight distribution in different design processes. In general, it is easy for the designers to interpret the scores in the context of different applications. When a design option received a low score for a criterion, it alerts the designer to have a close look and make attempts to improve it. When a criterion is adjusted to a lower weight in an assessment process, it would invite designers to ignore the criterion.}

\subsection{Orthogonal Names vs. Orthogonal Definitions}
\label{apx:Orthogonality}
We found that some conceptual overlapping is unavoidable, largely due to the broad interpretation of many words and the interrelated cognitive processes underpinning these concepts. It would be great if future research could propose a set of orthogonal names for different criteria.
Even if this were possible, it would take some time before the new names and definitions were to emerge and be used consistently in the literature. One may compare this desire to setting examination papers. It is usually not trivial to determine what is the optimal number of questions and whether the questions are orthogonal enough.

As an example of unavoidable overlap shown in Fig. \ref{fig:Criteria}, the concept of ``searchability'' (C5) in \cite{Chung:2016:CGF} is connected to several other concepts in \cite{Maguire:2012:TVCG,Borgo:2013:STAR,Chung:2016:CGF}. In general, a visual channel for encoding a data variable is expected to be distinguishable from the background and neighboring visual objects, though it may suffer from various shortcomings, e.g., being too small (Section \ref{sec:Geometry}), not having enough contrast with the background (Section \ref{sec:Colorimetry}), too much interference from the neighboring objects (Section \ref{sec:Separability}), lack of preattentiveness (Section \ref{sec:Balance}), etc. While any of these shortcomings can affect searchability, a visual channel, which does not suffer from these shortcomings, can still be difficult to search.

As discussed in Section \ref{sec:Searchability}, the searchability criterion cannot be replaced by or included in other criteria easily. At least, we have not found a suitable semantic definitions for other criteria in order for any of them to include all aspects of searchability. Meanwhile, if we were to combine all these criteria into a single criterion, it would be too complex to evaluate and too fluid to score consistently.

We should not see such semantic intricacy as a demerit of this proposed scheme but rather as a starting point for discussion by the community of the most appropriate and unambiguous terminology to use. Other alternative schemes may emerge in the future and they may be considered to be better by the community. Through the evolution of a language, a word can have several semantic meanings, and many words can have partially overlapping meanings. While we strive to optimize the boundary and minimize the overlapping among criteria, we are also constrained by the available words used to label these criteria. Hence in this work, we have made every effort to define each criterion precisely (see Section \ref{sec:Scheme}), to minimize the overlapping, and to find the most suitable word(s) to describe each criterion. Because of the constraints of the available words, having a set of orthogonal names for different criteria may not be possible. Meanwhile, the definitions of the criteria in Section \ref{sec:Scheme} are reasonably orthogonal.

\subsection{\reviseTVCG{Strategies for Defining the Five Levels of Each Criterion}}
\label{apx:Orthogonality}

\reviseTVCG{In the literature, Bertin \cite{Bertin:2011:book} assessed his criteria with two quality levels, i.e., ``yes'' or ``no''. Maguire et al. \cite{Maguire:2012:TVCG} used 5-level assessment for two criteria and 3-level assessment for one criterion, in addition to binary assessment for Bertin's criteria. They did not intend to combine these assessment scores numerically. Chung et al. \cite{Chung:2016:CGF} and Borgo et al. \cite{Borgo:2013:STAR} did not define assessment levels for their criteria. Nevertheless, they must have had at least two levels in mind, i.e., meeting or failing a criterion.}

\reviseTVCG{In general, most MCDA schemes in the literature used a 5-level or 10-level scoring system that is applied to all criteria in the same scheme consistently. While all VIS researchers and practitioners have been dealing with multiple design criteria in their work, using a MCDA scheme is not common. We thus propose to introduce a 5-level scoring system, which is a major step forward considering what is in the VIS literature. The VIS community is also familiar with 5-level scoring as major VIS conferences all use a 5-level scoring system for reviewing papers. Meanwhile, it may be premature to introduce a 10-level scoring system because the more levels an MCDA scheme has, the more precise specification is required. One can relate such a challenge to marking an essay with a marking scheme that has two levels (pass, fail), three levels (distinction, pass, fail), five levels (A, B, C, D, E), 11 levels (0 $\sim$ 10), or 101 levels (0 $\sim$ 100). The number of levels usually correlates with the level of preciseness requirement for a marking scheme and the level of difficulty in specifying such requirement.}

\reviseTVCG{We adopted a few relatively obvious approaches to the specification of the five levels:
\begin{itemize}
    \item A higher value is considered to be better. Hence 5 is the best and 1 is the worst.
    \item Level 3 is considered to be easily attainable by a relatively experienced VIS designer in most situations. There can be some exceptional circumstances, which will be discussed later in this Section.
    \item Level 4 is defined as the inbetweening point between Levels 3 and 5, and Level 2 is defined as the inbetweening point between Levels 1 and 3.
\end{itemize}
}

\reviseTVCG{We also adopted a few strategies that we consider to be realistic and sensible based on the current state of the art in terms of the theoretical understanding and available empirical data in the field of visualization. These include:}

\vspace{2mm}\noindent
\reviseTVCG{%
\textbf{Assuming that VIS knowledge is necessary.} Each scoring decision is expected to be made by a relatively experienced VIS designer, rather than a person with limited VIS design knowledge or a computer. Using such human knowledge should be considered as a positive aspect of the proposed MCDA scheme, rather than dismissing it as subjective or biased. In comparison with a design process without any scoring mechanism and specification of levels, the proposed MCDA scheme is less subjective and less biased. Meanwhile, our theoretical understanding about VIS designs is not ready for hand-crafting an algorithm to score each glyph design option, while there is not enough human-scoring data to train a machine-learning model to do so either. In the real-world, there are many MCDA schemes (e.g., building health-safety assessment, candidate selection assessment, and essay marking) for which expert knowledge has always been considered necessary in determining scores.}

\vspace{2mm}\noindent
\reviseTVCG{%
\textbf{Focusing on one or two key indicators to assess individual criterion.} When we have a \textbf{close} look at each criterion, it is not difficult for most of us to notice that its assessment could potentially involve the consideration of several factors. In other words, one could specify another MCDA scheme for assessing each criterion. This could go further into nested MCDA iterations in a hairsplitting manner.}

\reviseTVCG{\emph{Note: Such nested iterations are about breaking one criterion into multiple sub-criteria, and then breaking down each sub-criterion into multiple sub-sub-criteria. They are not the same as Type A, Type B, and Type C iterations as discussed in Section \ref{sec:Overview} as the latter break down the assessment of a whole glyph (Type D) into that of the encoding of individual variables (Type A) or their comparison (Type B and Type C).}}

\reviseTVCG{
It would cost so much time to complete a full round of assessment. We therefore focus the assessment of each criterion on one or two key indicators. This is somehow similar to setting examination questions for a course (e.g., calculus). At the high-level, the lecturer may set $N$ questions to cover $N$ topics of the taught courses (e.g., $N=10$, 10 marks each). Likely, most lecturers will try to ensure a good coverage (cf. the 12 criteria of the proposed MCDA scheme). At the level of each individual topic, the lecturer may focus on one aspect with one example (e.g., Q: $\int_0^\infty \frac{1}{a^2 + x^2}dx$, A: $\frac{\pi}{2a}, a \neq 0$). Based on the lecturer's experience, the assessment of $\int_0^\infty$ correlates with similar aspects (e.g., $\int_0^1$, $\int_{-\infty}^{\infty}$, etc.) The lecturer thus treats $\int_0^\infty$ as a key indicator. Meanwhile, there can be numerous examples of $f(x)$, and the lecturer can only select one or a few. This is also similar to the scenario that only a small number of examples and case studies could be provided in any paper.}

\reviseTVCG{
For example, there are many different types of color degeneration. The testing defined in Section \ref{sec:Colorimetry} involves two aspects, \emph{contrast} and \emph{brightness}. Many other aspects are not assessed directly, e.g., missing one of the RGB channels (a common error during overhead projection), different gamma settings of computer screens, and different environmental lighting conditions. Nevertheless, we can all appreciate some correlation between the two tested aspects and those not tested explicitly.}

\reviseTVCG{
While we focus on one or two key indicators for most criteria, the only exception is the first criterion, Typedness, which is assessed with four key indicators, i.e., Bertin’s
four kinds of perception (KOP), namely, associative, selective, ordered, and quantitative perception. This is partly because we do not wish to split Bertin's KOPs into different sub-criteria or select 1 $\sim$ 3 out of the four KOPs. Partly, and perhaps as a more significant reason, we anticipate most (if not all) VIS designers are familiar with Bertin's KOPs for the six visual channels in Table \ref{tab:BertinRating}, and they can make judgment about the typedness of these six visual channels quickly. In the worst-case scenario, having a quick glance at Table \ref{tab:BertinRating} does not take much time. As a glyph designer may potentially use many other visual channels, we also provide Table \ref{tab:ChannelRating} as provisional judgment decisions, which will need to be finalized by many empirical studies in the future. These two tables are further discussed in Appendix \ref{apx:Bertin}.}

\vspace{2mm}\noindent
\reviseTVCG{%
\textbf{Having one recommended solution is better than having none or waiting for reaching an agreement within the VIS community.} While most in the VIS community may agree the aforementioned approach to focus on one or two key indicators, it may not be easy to reach an agreement on choosing which key indicator for every criterion. In some cases, there are many options and we might have to wait for many years to reach an agreement. For example, for varying the contrast and brightness of an glyph image, there are many formulae in the literature in addition to the one given in Section \ref{sec:Colorimetry}. One may argue to use a non-linear formula or a local adjustment algorithm and start a new discussion about which non-linear formula or local algorithm. Even for the given linear formula, there can be many minor variations, such as replacing 259 with another number $> 255$ (avoiding division by 0) or replacing 128 with a ``better-defined'' middle gray. We decided to use 259 as there was a reasonable rationale in an online discussion \cite{Loch:2021:web}. We decided to use the more ``straightforward'' and ``obvious'' definition of middle gray, $128 \approx 255 \div 2$, because we consider that it is beyond the scope of this paper to debate about using which ``sophisticated'' definitions (e.g., ``18\% reflectance in visible light'' or the middle grey in CIEXYZ, CIELUV, CIELAB, or HSLuv color spaces).   
}

\reviseTVCG{%
We anticipate that discussions about selecting different key indicators for each criterion or different measuring formulae for each key indicator will continue as part of the VIS research. Such discussions should be encouraged and can be carried out through future research publications.   
}

\vspace{2mm}\noindent
\reviseTVCG{%
\textbf{Making level specification independent of the number of data variables.}
Most glyph designs in the VIS literature involve 10 or fewer data variables. Occasionally, there are more than 10 data variables (e.g., 20 in Duffy et al. \cite{Duffy:2015:TVCG}). The more data variables that a glyph has to encode, the more challenging for a design to be scored high against all criteria. We consider this is a norm. We do not wish for the MCDA scores to be used as an \emph{absolute} comparison across designs in different applications (specially in paper review processes). Consider two glyph designs X and Y. X involves many data variables as required in an application context, while Y involves only a few data variables. When X is scored lower than Y using the MCDA scheme, we should not consider that X is the worse design.}

\reviseTVCG{
Several criteria can easily be affected by the number of data variables. For example, the more data variables, the more difficult to achieve (i) high separability, (ii) good attention balance, (iii) high searchability, (iv) high learnability, and (v) high memorability. Some criteria may be indirectly affected. For example, discernability is recommended to be assessed for each data variable individually (i.e., Type A, direct assessment). However, with many data variables to be encoded, a designer may have to use smaller data objects or less powerful visual channels for some data variables. Nevertheless, when one obtains a Type D score through aggregation, the lower scores for some individual data variables may not impact on the Type D score noticeably.}

\reviseTVCG{
It may be helpful to note that the five levels of discernability are defined using percentage estimation across $n$ pairs of values (ranges). This is very different from across $n$ data variables. The latter would mean dependence on the number of data variables. In information theory, it is easy to distinguish the meanings of variable and value. Variables are alphabets and values are letters in an alphabet. Hence, using percentage estimation across $n$ pairs of values (ranges) does not contradict with this particular strategy.}

\reviseTVCG{
Some may wonder why not using absolute values, such as $\geq 40$ pairs, $40 \sim 39$ pairs, $20 \sim 29$, $10 \sim 19$ pairs, and $< 10$ pairs. The main reasons are:
\begin{itemize}
    \item The percentage-based specification is easy to define, learn, and memorize, while the counting-based specification has to consider the actual number $n$ (e.g., when $k = 5, n = 10$). For $k$ key values (ranges) in a data variable, there are $n = k(k-1)/2$ pairwise comparisons.
    \item When $n$ is a big number, it is quicker to estimate percentage than counting.
    \item Based on the authors' experience, it is common to over-specify the number $k$ for the key values (ranges) in each data variable during requirements analysis, as visualization is still commonly perceived as a tool for retrieving data, despite some empirical studies showing that for data retrieving tasks, visualization normally does not have any advantage over data tables \cite{Kanjanabose:2015:CGF}. Since visual channels in a glyph have very limited bandwidth, glyphs are in general worse than other forms of visualization for data retrieving tasks. We anticipate that this misconception will take many years to be adjusted, especially among non-VIS domain experts. Using percentage alleviates the problems caused by over-specification of the number $k$.     
\end{itemize}  
}

\begin{table}[th]
    \centering
    \caption{Bertin's rating of six visual channels. The cells with ``(can be)'' were rated as ``no'' by Bertin initially. A fair amount of evidence found in the literature has indicated that they ``can be'' meeting the requirements of the criteria concerned.}
    \begin{tabular}{@{}rcccc@{}}
        \textbf{Channel$\;$} & \textbf{Associative} & \textbf{Selective}
        & \textbf{Ordered} & \textbf{Quantitative}\\
        \hline
        Size        & no  & limited & yes & yes \\
        Brightness  & no  & yes & yes & yes \\
        Texture     & yes & yes & yes & no \\
        Color       & yes & yes & (can be) & (can be) \\
        Shape       & yes & (can be) & no & no \\
        Orientation & yes & yes & (can be) & (can be)\\
        \hline
    \end{tabular}
    \label{tab:BertinRating}
\end{table}

\begin{table}[th]
    \centering
    \caption{Tentative ratings for the channels listed in Chen et al. \cite{Chen:2014:book}. We used the term ``maybe'' to indicate our own uncertainty.}
    \begin{tabular}{@{}r@{\hspace{2mm}}c@{\hspace{2mm}}c@{\hspace{2mm}}c@{\hspace{2mm}}c@{}}
        \textbf{Channel$\quad$} & \textbf{Associative} & \textbf{Selective}
        & \textbf{Ordered} & \textbf{Quantitative}\\
        \hline
        \multicolumn{5}{c}{\emph{Geometric Channels:}}\\
        Size        & no  & limited & yes & yes \\
        Orientation & yes & yes & (can be) & (can be)\\
        Shape       & yes & (can be) & no & no \\
        Curvature   & no  & limited & yes & maybe \\
        Smoothness  & limited & limited & yes & maybe \\
        \hline
        \multicolumn{5}{c}{\emph{Optical Channels:}}\\
        Brightness  & no  & yes & yes & yes \\
        Color       & yes & yes & (can be) & (can be) \\
        Opacity     & no  & limited & yes & maybe \\
        Texture     & yes & yes & yes & no \\
        Shading     & limited & limited & yes & limited \\
        Halos       & limited & limited & yes & yes \\
        Shadow      & yes & yes & maybe & maybe \\
        Photo effects & limited & limited & maybe & maybe \\
        Implicit motion & limited & limited & maybe & maybe \\
        Explicit motion & yes & yes & yes & yes \\
        \hline
        \multicolumn{5}{c}{\emph{Relational Channels:}}\\
        Connection/edge & limited & limited & no & no \\
        Node & limited & limited & no & no \\
        Inside/outside & limited & limited & no & no \\
        Enclosure/Boundary & limited & limited & no & no \\
        Distance & no & limited & yes & yes \\
        Closure/opening & limited & limited & no & no \\
        Connectivity & yes & yes & maybe & maybe \\
        Partition & yes & yes & maybe & maybe \\
        Intersection/overlap & limited & limited & maybe & maybe \\
        Depth ordering & limited & limited & yes & maybe \\
        Hierarchy/level & limited & limited & yes & maybe \\
        Density/distribution & yes & yes & yes & no \\
        Convexity & limited & limited & no & no \\
        Continuity & limited & limited & maybe & no \\
        Genera & limited & limited & maybe & no \\
        Similarity & limited & limited & maybe & no \\
        Deformation & limited & limited & no & no \\
        \hline
        \multicolumn{5}{c}{\emph{Semantic Channels:}}\\
        Number & yes & yes & yes & yes \\
        Text & yes & yes & maybe & no \\
        Symbol/ideogram & yes & yes & maybe & no \\
        Sign/pictogram & yes & yes & maybe & no \\
        Isotype & yes & yes & maybe & no \\
        \hline
    \end{tabular}
    \label{tab:ChannelRating}
\end{table}

\section{Typedness and Bertin's Criteria}
\label{apx:Bertin}
Bertin's original criteria can be considered rateable using a ternary score (yes, limited, or no) based on Bertin's description as shown in Table \ref{tab:BertinRating}. Some cells in the table were rated as ``no'' by Bertin \cite{Bertin:2011:book}. However, a fair amount of evidence found in the literature has indicated that they ``can be'' considered to meet the requirements of the criteria concerned. For example, colors are widely used in heatmap visualization, and orientation can be used for encoding angles and data values suitable for a clock metaphor. 

In the context of Section \ref{sec:Typedness}, we can consider a ``yes'' to be appropriate for a kind of perception (KOP), and a ``no'' to be inappropriate. The terms ``limited'' and ``(can be)'' indicate that the assessment needs to take application-specific requirements into account.

There are many visual channels that are yet to be assessed using Bertin's four criteria. For example, Chen et al. listed more than 30 visual channels \cite{Chen:2014:book}. Assessing all these channels is beyond the scope of this paper. In Table \ref{tab:ChannelRating}, we provide tentative ratings for these channels to aid the users of our MCDA scheme.

\section{Tried-and-Tested Workflows}
\label{apx:Workflow}
In this appendix, we describe 12 workflows for assessing the 12 criteria, respectively, outlined in the main body of the paper. In November 2021, we began to abstract our experience in designing glyphs into a set of criteria for analyzing the trade-offs among different criteria. We drew our experience from the previous practice in designing glyphs for different applications (including those reported in \cite{botchen2008action,Maguire:2012:TVCG,Legg:2012:CGF,Borgo:2013:STAR,Duffy:2015:TVCG,chung2015glyph,Chung:2016:CGF,legg2016glyph}) as well as our recent experience of designing glyphs for gait analysis and music analysis in undergraduate and doctoral projects. The workflows reported here reflect our experience of using the 12 criteria in comparing different glyphs designs since December 2022 when the first version of the 12 criteria was completed (archived as arXiv:2303.08554 in March 2023).  
We believe that the proposed criteria and the accompanying workflows will be evaluated and improved by the VIS community after the work has been tested, critiqued, and adopted or adapted by VIS practitioners. This will take some time (e.g., the first paper that reported the adoption of the nested model was published four years after the paper on the nested model). We hope that at some stage, a set of recommended workflows will emerge in the VIS community.

\reviseTVCG{``Multi-criteria decision analysis'' is a standard term for a methodology in social sciences. The term ``decision'' can be interpreted in many ways relevant to glyph designs, e.g.,
\begin{itemize}
    \item \textbf{Selective decision} --- Given $N$ design options at an intermediate design stage, the designer selects $K < N$ designs to be put forward to the next meetings with domain experts. Based on our experience of meeting domain experts for discussing design options, discussing 4 $\sim$ 8 design options seems to be relatively effective. Domain experts often prefer to make comments such as ``I like this better than that.'' Having too many design options may overwhelm the domain experts, especially with many similarly-looking designs, while having too few may hinder domain experts' ability to compare and often leads to a hesitant positive answer ``it seems OK''.
    \item \textbf{Judgment decision} --- Given a new design proposed by one or a few designers, the designers themselves apply the scheme to the design to identify strengths and weaknesses of the design. Cognitively, the score sheets used for the MCDA enable external memorization for the judgment on different design criteria, where externalized information facilitates more effective communication among different designers.
    \item \textbf{Initial Judgment decision} --- At an early design stage, we also found that comparing each proposed design with a baseline design (e.g., $n$ variables are encoded as $n$ bars around a circle) quickly using the 12 criteria is also highly effective for filtering out designs that overly in favor of some criteria but have major issues with other criteria. A few undergraduate students working on glyph-based visualization as their final year projects used this approach to learn and appreciate different design criteria.
\end{itemize}}

\reviseTVCG{It is possible that after the MCDA scheme has been deployed, many VIS designers may use the MCDA scheme to make other types of decisions.}

\reviseTVCG{In many design processes, there may be more than one designer or more than one assessor. As described in Section \ref{sec:Evaluation}, we have experienced two mechanisms for deriving the final scores for a design option when there are multiple assessors. It is possible that after the MCDA scheme has been deployed, many VIS designers may report their experience of using these two mechanisms and propose new mechanisms.}

In the next 12 subsections, we outline 12 workflows for the 12 criteria respectively. Following these, we summarize our general observations about these workflows in the final subsection. In these subsections, we refer to the person who assesses different glyph designs as a ``rater''. A rater may be the designer who wishes to compare a set of provisional designs to filter out a small set for further discussions with the potential users, or be a relatively independent VIS specialist for assessing different designs produced by others (e.g., in a training course).

Before diving into the details of each criterion, we have a few general recommendations:
\begin{enumerate}
    \item[R1.] It is important to read the definition of each criterion in Section \ref{sec:Scheme} carefully, rather than relying on a personal interpretation of the name of the criterion. As discussed in Appendix \ref{apx:Orthogonality}, there may not always be appropriate words in the dictionary to convey the definitions of the criteria in an orthogonal manner.
    \item[R2.] When considering a glyph design, a rater should ideally see a few variable settings for the same design. For example, if a glyph encodes two data variables $d_1$ and $d_2$ where $d_1 \in \{1, 2, 3, 4, 5\}$  and $d_2 \in \{a, b, c, d\}$. The rater may observe a few instances, such as five glyphs for $(1, a)$, $(1, d)$, $(5, a)$, $(5, d)$, and $(3, b)$.
    \item[R3.] When there are a number of data variables and each variable has a number of valid values, there will be too many instances for a rater to observe. One important practice is for the rater to imagine what a glyph may look like in other variable settings that are not available to view. This is mostly a practice issue rather than a skill issue. Many visualization researchers and practitioners are often in a situation where one is given some data, and is asked to produce a visualization for the given data. On the other hand, when one is designing a system for handling dynamic data, one has to imagine different scenarios of the data that has not yet become available. When we design and evaluate glyphs, we must get into the habit of anticipating different scenarios, i.e., different variable settings.
\end{enumerate}


\subsection{Typedness}
\label{apx:Typedness}
In the process for designing a glyph to encode a small number of data variables, a rater may wish to assess each design directly using a Type D score directly. However, when the number increases, the rater could easily overlook some variables. By listing and scoring all variables to produce Type A scores first, e.g., using a spreadsheet, the explicit scoring sheet serves as a form of external memorization, which also happens to be a merit of visualization. The basic steps are:

\begin{enumerate}
    \item[1.] Identify $n$ data variables to be encoded, together with the visual identities of these variables. For the time being, we consider the identities as the valid values of a single nominal variable. Later, we will discuss these identities further.
    \item[2.] For each data variable, determine its type (i.e., nominal, ordinal, ratio, or interval).
    \item[3.] For each data variable, determine the KOPs (Bertin's kinds of perception) required by the variable, resulting in applicable KOPs (AKOPs).
    \item[4.] For each data variable, check if the visual encoding is appropriate, and assign a score based on the five levels defined in Section \ref{sec:Typedness}.
    \item[...] Repeat steps 2$\sim$4 for all $n+1$ variables.
    \item[fin.] Derive a type D score by calculating the average of the $n+1$ Type A scores. Here ``fin.'' is short for ``finally'', and it is also used in the other workflows in this appendix.
\end{enumerate}

Table \ref{tab:BertinKOP} lists Bertin's four kinds of perception (KOP) and their applicability to four types of variables \cite{bertin1983semiology}. 

\begin{table}[th]
    \centering
    \caption{Bertin's four kinds of perception (KOP).}
    \begin{tabular}{ccccc}
        \textbf{KOP} & \textbf{Nominal} & \textbf{Ordinal}
        & \textbf{Interval} & \textbf{Ratio}\\
        \hline
        Associative  & required  & required & required & required \\
        Selective    & required  & required & required & required \\
        Ordered      & not req. & required & required & required \\
        Quantitative & not req. & not req. & required & required \\
        \hline
    \end{tabular}
    \label{tab:BertinKOP}
\end{table}

In many existing research papers on glyphs, the authors often focused the discussions on the variables of the data concerned, while the identities of these variables are commonly not discussed as a variable although they are always encoded one way or another. For example, in Design A by Maguire et al. \cite{Maguire:2012:TVCG} in Table \ref{tab:CaseStudy1}, the data variables S0, S6, S2 and S5 are identifiable by the locations of their visual channels as well as other visual variations, e.g., the outside shape for S0, filling color and shape for S6, the circular line type (always grey) for S2, and inside shape (always orange) for S2. 

Theoretically, given $n$ data variables, we can treat these identities as the $n$ valid values of a single nominal variable, which is encoded using multiple visual channels. We can thus assess it as a single variable. Alternatively we can break it down to $n$ identity variables and each identity variable is a binary variable, representing whether the corresponding data variable is visually distinguishable from others. Note that one should not confuse that the identity variable for a data variable with the associative and selective KOPs for the data variable. As shown in Table \ref{tab:BertinRating}, Bertin proposed to assess the KOPs of a variable (i.e., its visual encoding) independently from other variables. The assessment of an identity variable or variables has to consider the identity encoding for all $n$ data variables.  

\subsection{Discernability}
\label{apx:Discernability}
Similar to the Typedness criterion, we recommend to assess each variable individually and then take the average of the obtained Type A scores to produce a Type D score.

It is important to stress that the goal of encoding a numerical variable is not for us to read the number from the visual encoding. As already evidenced in empirical studies (e.g., Kanjanabose et al. 2015, 10.1111/cgf.12638), for such a goal, data tables usually allow humans to perform their tasks more accurately and quickly than visual objects that encode numbers. As the visual objects in a glyph are much smaller than those in most visualization plots, over-focusing on the goal of reading numbers can only be a distraction from the real uses of glyphs.

Previous work on glyph-based visualization typically features applications where many glyphs are displayed in the same visualization (e.g., \cite{Legg:2012:CGF,Maguire:2012:TVCG,Duffy:2015:TVCG,chung2015glyph,legg2016glyph}), and the visualization tasks concerned include observing temporal changes in a group of glyphs, comparing different glyphs, identifying clusters and anomalies, and so on. Of course, to perform these tasks, one needs to be able to discern encoded values in each variable at a certain resolution. A glyph design ideally could achieve an optimal trade-off among (i) keeping its size reasonably small, (ii) enabling high discernability, and (iii) enabling the efficient performance of the tasks to be supported by glyph-based visualization. This trade-off is often characterized by the balance between a design with a large glyph and a design allowing many glyphs to be displayed simultaneously. Such trade-offs are encapsulated by the information-theoretical analysis of the cost-benefit of visualization (Chen and Golan 2016, 10.1109/TVCG.2015.2513410).

One important step in assessing the discernability of a variable is to determine the data values or data ranges that need to be discernable. For example,
\begin{itemize}
    \item \emph{Nominal and ordinal variables} -- If such a variable has $n$ valid values, must all $n$ values be discernable? When $n$ is a relatively large number (e.g., $n > 8$), would grouping be helpful or problematic for tasks such as observing temporal patterns or identifying anomalies?
    \vspace{-1mm}
    \item \emph{Interval and ratio variables} -- Such a variable usually has a large number of valid values within the range, or even an infinite number of values theoretically. Almost all visualization representations have to deal with the discernability issues, e.g., in a heat-map, is a pixel color distinguishable from another; in a line plot, is a point on a time series discernable from another point a few seconds later. In glyph-based visualization, typically the valid values are divided into $k$ sub-ranges, with different critical values falling into different sub-ranges. Even when continuous visual mapping is used (e.g., continuous line length or color-mapping), one needs to determine what level of perceptual discernability is required.  
\end{itemize}

When a data variable is encoded using multiple visual channels, the scoring focuses on the most discernable visual channel. For example, imagine that one needs to encode a real variable describing music volume. One decides that six levels are sufficient in a glyph design. One may encode six levels as five stacked color strips, i.e., no strip for 0, 1 strip for level 1, etc. The visual channels used include size of the colored area, aspect ratio of the colored area, location of the top strip, and strip counting. Among these visual channels, strip counting is most reliable in terms of discernability.

The main steps for assessing discernability are therefore as follows:

\begin{enumerate}
    \item[1.] Identify $n$ data variables to be encoded, together with the visual identities of these variables.
    \item[2.] For each data variable, determine its $k$ values (or $k$ groups of its values) that must be discernable. 
    \item[3.] For each data variable, compare all $k(k-1)/2$ pairs of values (or groups, ranges) to see if they can be differentiated at ease, and assign a score based on the five levels defined in Section \ref{sec:Discernability}.
    \item[...] Repeat steps 2$\sim$3 for all $n+1$ variables.
    \item[fin.] Derive a type D score by calculating the average of the $n+1$ Type A scores.
\end{enumerate}


\subsection{Intuitiveness}
\label{apx:Intuitiveness}
When we were formulating the MCDA scheme, we had extensive discussions on three related criteria, Intuitiveness, Learnability, and Memorability. Some aspects, such as the total number of data variables, which affect Learnability and Memorability, should ideally be assessed ``globally'' for the whole glyph. Some aspects, such as domain-specific convention for visual mapping (e.g., mapping music volume to height), should ideally be assessed ``locally'' for individual variables. Some aspects, such as visual metaphors, may need to be assessed both ``locally'' (e.g., the metaphoric visual encoding for a data variable) and ``globally'' (e.g., a metaphoric layout). If the ``local'' and ``global'' were mixed into the same criterion, it would be challenging to disentangle different aspects. On balance, we found that separating ``local'' and ``global'' assessment helps structure these aspects into different criteria, so the rater can be more focused in assessing each criterion. The main steps for assessing Intuitiveness are as follows:

\begin{enumerate}
    \item[1.] Identify $n$ data variables to be encoded, together with the visual identities of these variables.
    \item[2.] For each data variable, list the domain-specific conventions for encoding this variable.
    \item[3.] Determine if the proposed visual encoding features an additional visual metaphor. 
    \item[4.] For each data variable, consider the quality of the visual encoding in terms of domain-specific conventions (if any) and visual metaphor (if any), and assign a score based on the five levels defined in Section \ref{sec:Intuitiveness}.
    \item[...] Repeat steps 2$\sim$4 for all $n+1$ variables.
    \item[fin.] Derive a type D score by calculating the average of the $n+1$ Type A scores.
\end{enumerate}

\reviseTVCG{Consider the example in Fig. \ref{fig:Intuitiveness}, which was one of seven variables in the application reported by Maguire et al. \cite{Maguire:2012:TVCG}. It is variable $S5$, and it has seven categorical values, namely
(C1) input molecular part,
(C2) input cellular component,
(C3) input cell,
(C4) input tissue,
(C5) input organ,
(C6) input organism,
(C7) input population.
In the application concerned, these are seven levels of granularity associated with material entities in a biological workflow. The domain experts need to inspect many workflows routinely and there were tens of thousands of workflows in a data repository managed by these domain experts when the paper \cite{Maguire:2012:TVCG} was written. The designs in Fig. \ref{fig:Intuitiveness} and some other designs in \cite{Maguire:2012:TVCG} were all considered by Maguire et al., and many more designs were drawn on whiteboards at that time.
The scores in Fig. \ref{fig:Intuitiveness} were given by the authors of this work after one co-author recalled the discussions by  Maguire et al. when their work was carried out.}

\reviseTVCG{The domain-specific convention (DC) was considered as the seven levels, i.e., 1, 2, $\ldots$, 7. There was no standard visual representation for these seven levels, and photo-realistic images or fairly realistic sketches were often used as illustrations. As the first three rows all have the numbers 1, 2, $\ldots$, 7 encoded, they are labeled with cnDC (consistent with the DC), while the bottom three rows are labeled with inDC (inconsistent with the DC). The 2nd and 5th rows do not have any additional visual metaphor (AM), and we labeled them as noAM.}

\reviseTVCG{The 1st and 4th rows were considered by domain-expert co-authors of \cite{Maguire:2012:TVCG} as appropriate, and we labeled them as apAM. There were many versions of these, abstracted from some photo-realistic images or fairly realistic sketches used for illustrations. The design in the 5th row was first obtained, and the design in the 1st row was then derived by ``injecting'' numbers into the shapes in the 5th row.}

\reviseTVCG{The 3rd row attempted to create a visual metaphor from the numbers. It was considered too difficult to perceive the visual metaphors. Hence we labeled it inAM. Some visual metaphors (e.g., the smiley face and the cloud) in the 6th row were considered as inappropriate. The small and large circles were also considered to be metaphorically too close to each other. We thus labeled the design option as inAM. Note that the label inAM is assigned to a variable, rather than an individual categorical value. If one categorical value is encoded inappropriately, the encoding of the whole variable is considered as inAM.   
}

One reviewer of this paper drew our attention to a broad discussion on the assessment of intuitiveness (e.g., \cite{Naumann:2007:book,Reinhardt:2024:IJHCF}). We hope that future research in this broad context will help improve the intuitiveness assessment in the relatively small context of glyph design.  

\subsection{Invariance: Geometry}
\label{apx:Geometry}
As discussed in Section \ref{sec:Geometry}, we recommend to evaluate this criterion with a Type D score directly as it is more efficient and effective to scale a glyph design as a whole object. Likely different raters will work out their own ways to scale a glyph being assessed. In the future, a software tool for glyph design will likely be equipped with scaling functions for evaluating geometrical invariance. With the current software provision, a glyph designer may draw glyphs by programming or using a drawing tool, such as Adobe Illustrator, Microsoft PowerPoint, or Apple Keynote. When glyphs are created by writing programs, the designer can easily draw glyphs with different scaling factors. When glyphs are created using drawing tools, one can also scale them using such tools. Using Microsoft PowerPoint as an example, after drawing a glyph, one may scale it with the following steps:

\begin{enumerate}
    \item[1.] Group all graphical elements together to form a single object.
    \item[2.] Copy the single object into the clipboard, and paste it as a picture (e.g., Enhanced Metafile or PNG). Make five ``picture'' copies in the same format. 
    \item[3.] Open the ``Drawing'' panel and the ``Size'' sub-panel, and make sure that ``Lock aspect ratio'' is on.
    \item[4a] If the glyph design is a circular design, change the widths of the five pictures to 4.37cm, 3.49cm, 2.62cm, 1.75cm, and 0.87cm respectively. One should see five glyphs of different sizes similar to the middle row of Fig. \ref{fig:Invariance}. See Section \ref{sec:Geometry} for the definitions of these values.
    \item[4b] If the glyph design is a rectangular design, determine if the width or height will be used to control the scaling. One normally selects whichever is shorter. Assuming the width is selected, change the widths of the five pictures to 3.09cm, 2.47cm, 1.85cm, 1.23cm, and 0.62cm respectively. One should see five glyphs of different sizes similar to the bottom row of Fig. \ref{fig:Invariance}. See Section \ref{sec:Geometry} for the definitions of these values.
    \item[5.] Align the five resized glyphs on the same slide.
    \item[6.] Draw a square of 4cm$\times$4cm (digilally) on the same slide, and zoom-in or out the slide to ensure that the square measures 4cm$\times$4cm (physically) on the monitor.
    \item[7.] Adjust the viewing distance from the eyes to the monitor to be 50cm.
    \item[fin.] Observe the glyphs carefully and assign a score based on the five levels defined in Section \ref{sec:Geometry}.
\end{enumerate}

\subsection{Invariance: Colorimetry}
\label{apx:Colorimetry}
Most imaging processing tools and some drawing tools provide facilities to adjust the brightness and contrast of an image. However, they do not necessarily using the same formula. Hence there will be some inconsistency when different raters are using different tools. Even if all raters were using the same tool, the perceptual judgment among different people would not be consistent. Different monitors may also affect the perception. Ideally, there would be glyph design tools equipped with the same formula for adjusting the brightness and contrast of a glyph, and for assessing the levels of the chromatic and achromatic invariance algorithmically based on appropriate computer monitor data and human perception data collected from extensive empirical studies. It will likely take years or decades to achieve this. Until then, raters can make use of existing tools to evaluate this criterion. For example, one may use Microsoft PowerPoint by following a sequence of steps similar to those in Appendix \ref{apx:Geometry}.

\begin{enumerate}
    \item[1.] Group all graphical elements together to form a single object.
    \item[2.] Copy the single object into the clipboard, and paste it as a picture (e.g., Enhanced Metafile or PNG). Make 20 ``picture'' copies in the same format, and organize them into a $4 \times 5$ grid, i.e., four rows by five columns. 
    \item[3.] Open the ``Drawing'' panel and the ``Picture'' sub-panel.
    \item[4.] For the first column of glyph pictures, make no changes. For the second column of four glyphs, change the brightness ($B$) and contrast ($C$) as:
    \begin{itemize}
        \item $[B+10\% B,\; C+10\% C]$,
        \item $[B+10\% B,\; C-10\% C]$,
        \item $[B-10\% B,\; C+10\% C]$,
        \item $[B-10\% B,\; C-10\% C]$ 
    \end{itemize}
    For the third column, perform similar adjustment by replacing the previous $10\%$ with $20\%$. For the fourth column, perform similar adjustment by replacing the previous $20\%$ with $30\%$. For the fifth column, perform similar adjustment by replacing the previous $30\%$ with $40\%$.     
    \item[fin.] Compare the first column with each of the other four columns and assign a score based on the five levels defined in Section \ref{sec:Colorimetry}.
\end{enumerate}

\subsection{Composition: Separability}
\label{apx:Separability}
As discussed in Section \ref{sec:Separability}, obtaining $n(n-1)$ pairwise Type B scores $s_\text{int}(\lambda_i, \lambda_j)$ would be time-consuming. Note that we consider that $(\lambda_i, \lambda_j)$ is a directional pair, as it is common that $s_\text{int}(\lambda_i, \lambda_j) \neq s_\text{int}(\lambda_j, \lambda_i)$. It is therefore more cost-effective to estimate Type C scores or a Type D score. Let us first describe how to estimate Type C scores $S_\text{int}(\lambda_i)$ that is defined by a $\max()$ formula, and we follow this by describing the methods for estimating $\max_\text{int}$ and $\text{avg}_\text{int}$, which determine a Type D score.

$\blacktriangleright$ Estimating $S_\text{int}(\lambda_i)$ --- Given $n$ visual channels in a glyph, $\lambda_1, \lambda_2, \ldots, \lambda_n$, $\max_\text{int}(\lambda_i)$ defines the maximal level of interference $\lambda_i$ that could receive from other visual channels. As $s_\text{int}(\lambda_i, \lambda_j)$ may take one of four possible values 1 (major), 0.1 (medium), 0.01 (minor), and 0 (none), $S_\text{int}(\lambda_i)$ can only take one of these four values. Hence to estimate $S_\text{int}(\lambda_i)$, the rater needs to identify a visual channel $\lambda_j$ ($j \neq i$) that may give most undesirable interference to $\lambda_i$. The rater needs to focus only on the worst case scenario, which is cognitively less demanding than, e.g., taking the average of all Type B scores for $s_\text{int}(\lambda_i, \lambda_j), j \neq i \land j = 1, 2, \ldots, n$.

$\blacktriangleright$ Estimating $\max_\text{int}$ --- For the whole glyph, the score $\max_\text{int}$ is also defined by a $\max()$ formula. Hence the rater only needs to consider the worst case scenario among $S_\text{int}(\lambda_1), S_\text{int}(\lambda_2), \ldots, S_\text{int}(\lambda_n)$. Hence obtaining $\max_\text{int}$ is not cognitively demanding. 

$\blacktriangleright$ Estimating $\text{avg}_\text{int}$ --- In Section \ref{sec:Separability}, $\text{avg}_\text{int}$ is defined with an average formula, which seems to be cognitively demanding to estimate if one were to use mental calculation. However, because Type C scores have only four levels, i.e., 1 (major), 0.1 (medium), 0.01 (minor), and 0 (none), we can use counting primarily to estimate $\text{avg}_\text{int}$ as shown in Algorithm \ref{alg:Separability}. This is not a complex algorithm because lines 1-8 are very similar to lines 10-17. There are only three pathways for running the algorithm, (i) lines 1-8, (i) lines 1, 2, 9-17, and (ii) lines 1, 2, 9-11, 18-20. Hence it is possible to run through this counting process in one's mind.

\begin{algorithm}[t]
  Count the number of 1's among $S_\text{int}(\lambda_1), S_\text{int}(\lambda_2), \ldots, S_\text{int}(\lambda_n)$, which is denoted as $K_a$\;
  \eIf{$K_a > 0$}{
    $T := K_a$\;
    Count the number of 0.1's among $S_\text{int}(\lambda_1), S_\text{int}(\lambda_2),$ $\ldots, S_\text{int}(\lambda_n)$, which is denoted as $K_b$\;
    \uIf{$K_b \in [5, 14]$}{
      $T := T + 1$\;
    } \uElseIf{$K_b \in [15, 24]$}{
      $T := T + 2$\;
    }
    \tcp{\small add more \textbf{else if} when there are 25+ channels}
  } {
  Count the number of 0.1's among $S_\text{int}(\lambda_1), S_\text{int}(\lambda_2),$ $\ldots, S_\text{int}(\lambda_n)$, which is denoted as $K_b$\;
  \eIf{$K_b > 0$}{
    $T := 0.1 K_b$\;
    Count the number of 0.01's among $S_\text{int}(\lambda_1), S_\text{int}(\lambda_2),$ $\ldots, S_\text{int}(\lambda_n)$, which is denoted as $K_c$\;
    \uIf{$K_c \in [5, 14]$}{
      $T := T + 0.1$\;
    } \uElseIf{$K_c \in [15, 24]$}{
      $T := T + 0.2$\;
    }
    \tcp{\small add more \textbf{else if} when there are 25+ channels}
  } {
    Count the number of 0.01's among $S_\text{int}(\lambda_1), S_\text{int}(\lambda_2),$ $\ldots, S_\text{int}(\lambda_n)$, which is denoted as $K_c$\;
    $T := 0.01 K_c$\;
    }
  }
  $\text{avg}_\text{int} := T / n$\;
\caption{Mental steps for estimating $\text{avg}_\text{int}$}
\label{alg:Separability}
\end{algorithm}

We describe below the steps for estimating and writing down Type C scores first, then estimating $\max_\text{int}$ and $\text{avg}_\text{int}$, and finally working out the Type D score.

\begin{enumerate}
    \item[1.] Select one visual channel $\lambda_i\; (i = 1, 2, \ldots, n)$ for assessment.
    \item[2.] Examine if $\lambda_i$ may potentially receive interference from any of the other channels. Estimate and write down $S_\text{int}(\lambda_i)$ as described earlier.
    \item[...] Repeat steps 1$\sim$2 for all $n$ variables.
    \item[fin.] Estimate $\max_\text{int}$ and $\text{avg}_\text{int}$ as described earlier. Assign a score based on the five levels defined in Section \ref{sec:Separability}.
\end{enumerate}

For an experience rater, because $\max_\text{int}$ and $\text{avg}_\text{int}$ can be estimated mentally, the rater can assign a Type D score directly. 


\subsection{Composition: Comparability}
\label{apx:Comparability}
Although Type B and Type C scores may seem applicable to this criterion, it is not as complex as it appears, because we only need to consider those pairs of data variables that need to be compared within a glyph. Given $n$ data variables, we can divide them into $k$ groups, each with $n_i\; (i=1,2,\ldots,k)$ number of variables. Therefore, $n_1 \geq 0$, $n_i > 0\; (i=2, \ldots, k)$, and $ n = n_1 + n_2 + \ldots + n_k$. Here the first group is designated as the non-comparable group, i.e., there is no requirement to compare the $n_1$ data variables in the first group. In many applications, $k$ is usually a small number. When $k = 1$, there is no need to compare any of the $n$ data variables. In this case, we do not assign any Type D score to this criterion, which will not contribute to the overall weighted average.

When $k > 1$, one needs examine $M$ pairwise relationships, where
\[
    M = \frac{n_2 (n_2 - 1) }{2} + \frac{n_3 (n_3 - 1) }{2} + \ldots + \frac{n_k (n_k - 1) }{2}
\]
\noindent As $k$ is usually a small number, the actual situations are not as complex as the impression given by the formula. The following two scenarios are relatively common.
\begin{itemize}
    \item[a.] Most data variables in a glyph need to be compared, e.g., a glyph encodes the average incomes of $n_2$ social groups. Most glyph designers are expected to notice the need to compare these $n_2$ variables, and likely they would have used the same type of visual channels to encode these variables. The rater's role is to check for outliers.
    \item[b.] There are a few small groups of data variables that need to be compared, e.g., in Duffy et al. \cite{Duffy:2015:TVCG}, there are two such groups, with two and three data variables respectively. Therefore $k=3, n=20, n_1 = 15, n_2 = 2, n_3 = 3$. The rater would need to assess $M=4$ pairwise relationships. This is not a big number.
\end{itemize}

We will describe the workflow below as if we use Type B scores. In practice, with the above common scenarios (a) and (b), the rater can easily obtain a Type D score directly.   

\begin{enumerate}
    \item[1.] Identify all $M$ pairs of comparable relationships that need to be assessed.
    \item[2.] For each pairwise relationship, determine if there is a \emph{Major}, \emph{Medium}, and \emph{Minor} obstacle to hinder the comparability.
    \item[...] Repeat step 2 for all $M$ pairs.
    \item[fin.] Transform the $M$ assessment to a score based on the five levels defined in Section \ref{sec:Comparability}.
\end{enumerate}


\subsection{Attention: Importance}
\label{apx:Importance}
The mathematical definition for this criterion in Section \ref{sec:Importance} is for ensuring long-term consistency of the definition itself. In the short-term, the community is still waiting for more empirical studies that can inform us about the amount of attention that different visual channels may attract. As shown in Appendix \ref{apx:Typedness}, many visual channels have not yet been adequately studied.

Nevertheless, we can conduct the assessment at a ``lower resolution'' relatively easily. If we consider only two levels of importance and two levels of attention, we can place $n$ data variables into $2 \times 2$ boxes:
\begin{center}
\begin{tabular}{|c|c|c|}
    \hline
    & \textbf{1. less important} & \textbf{2. more important} \\
    \hline
    \textbf{1. less preattentive} & $n_{1,1}$ variables & $n_{1,2}$ variables \\
    \hline
    \textbf{2. more preattentive} & $n_{2,1}$ variables & $n_{2,2}$ variables \\
    \hline
\end{tabular}
\end{center}

\noindent Let $n_{1,1}$, $n_{1,2}$, $n_{2,1}$, $n_{2,2}$ be the numbers of data variables in these four boxes. We assign $\iota = 1$ to the ``less important'' category and $\iota = 2$ to the ``more important'' category. Similarly, we assign $\alpha = 1$ to the ``less preattentive'' category and $\alpha = 2$ to the ``more preattentive'' category.   We can actually compute the Pearson correlation index as:

\begin{align*}
    \overline{\iota} &= \frac{n_{1,1} + n_{2,1} + 2 n_{1,2} + 2 n_{2,2}}{n}\\
    \overline{\alpha} &= \frac{n_{1,1} + n_{1,2} + 2 n_{2,1} + 2 n_{2,2}}{n}\\
    A &= n_{1,1}(1-\overline{\iota})(1-\overline{\alpha}) +
    n_{1,2}(2-\overline{\iota})(1-\overline{\alpha})\\
    &+ n_{2,1}(1-\overline{\iota})(2-\overline{\alpha}) +
    n_{2,2}(2-\overline{\iota})(2-\overline{\alpha})\\
    B_\iota &= (n_{1,1} + n_{2,1})(1-\overline{\iota})^2 +
         (n_{1,2} + n_{2,2})(2-\overline{\iota})^2\\
    B_\alpha &= (n_{1,1} + n_{1,2})(1-\overline{\alpha})^2 +
         (n_{2,1} + n_{2,2})(2-\overline{\alpha})^2\\
    C &= \frac{A}{\sqrt{B_\iota B_\alpha}}
\end{align*}

A rater can simply make a standard spreadsheet for calculating $C$ with four input values $n_{1,1}$, $n_{1,2}$, $n_{2,1}$, $n_{2,2}$. Similar, if a rater wishes to use three levels for both importance and attention, the above procedure can be adapted for $3 \times 3$ boxes and a sightly different spreadsheet for calculating $C$. The following steps describe a procedures for generalized $k \times k$ boxes.

\begin{enumerate}
    \item[1.] Place each data variable into one of the $k \times k$ boxes.
    \item[2.] Count the number of data variables in each box.
    \item[3.] Use a spreadsheet to determine the correlation index $C$.
    \item[fin.] Map the value of $C$ to one of the five levels defined in Section \ref{sec:Importance}.
\end{enumerate}

\subsection{Attention: Balance}
\label{apx:Balance}
To assess this criterion, a rater needs to explore all data variables and all possible values that each variable may take, considering the attention that the variable may attract through its visual channel(s). When a data variable is encoded using multiple visual channels, focus on the most preattentive visual channel and the combined preattentive quality, rather than the visual channel with the weakest preattentive quality. As the rater may be presented with a limited number of glyph instances, the rater needs to imagine the preattentive quality of those glyphs that encode other values but are not presented. As long as a data variable can take a certain value, with which the data variable attracts an inadequate amount of attention, the rater should consider that this is a case of weak attention. In other words, the rater should consider the worst case scenario (rather than the average case) for a variable. The steps for assessing this criterion are relatively straightforward. 


\begin{enumerate}
    \item[1.] Scan every data variable to see if it receives weak attention, i.e., it cannot attract an inadequate amount of attention.
    \item[2.] Count the number of such data variables.
    \item[fin.] Map the counted number to one of the five levels defined in Section \ref{sec:Balance}.
\end{enumerate}

\subsection{Searchability}
\label{apx:Searchability}
As discussed in Section \ref{sec:Searchability} and Appendix \ref{apx:Orthogonality}, this criterion cannot be replaced by or included in other criteria easily, though its name seems to suggest some overlapping. Nevertheless, its definition separates this criteron from others. As defined in Section \ref{sec:Searchability}, the searchability criterion \emph{assesses the desirable property that the visual channel(s) for each data variable can be recognized easily among others after a viewer has learned and remembered the encoding scheme.} 

The notion of cognitive load is well-known in the fields of visualization and psychology, and is critical to understand how visualization works and how to optimize a visual design. However, the measurement of cognitive load has not reached the level that we can easily quantify the cognitive load of an action or task in a visualization process. Likely it will take decades to improve the current situation, possibly through conducting a huge number of empirical studies to collect data and/or developing more efficient and effective techniques for measuring cognitive load.

Nevertheless, the scheme requires a rater to estimate cognitive load but at only three levels, \emph{low}, \emph{medium}, and \emph{high}, as defined in Section \ref{sec:Searchability}. A rater can become used to the practice of estimating cognitive load at these three levels. Equipped with the skill for estimating cognitive load, a rater can produce a Type D score with the following steps. 

\begin{enumerate}
    \item[1.] Scan every data variable to see if searching it will likely demand a \emph{low}, \emph{medium}, or \emph{high} amount of cognitive load.
    \item[2.] Count the numbers of such data variables in the categories of \emph{low}, \emph{medium}, and \emph{high} respectively.
    \item[fin.] Map the counted numbers to one of the five levels defined in Section \ref{sec:Searchability}.
\end{enumerate}



\subsection{Learnability}
\label{apx:Learnability}
An experienced rater will likely be able to anticipate the time required to learn a glyph design. Alternatively, a rater or designer can organize meetings with potential users to observe how quickly potential users can grasp the visual encoding scheme and how easily they may be confused by the semantic meanings of the visual encoding. As it is common for glyph designers to hold meetings with potential users for requirement analysis and evaluation, assessing learnability can be part of such meetings.

For an independent rater who may not have the access to the potential users, the rater can gather learnability information from the glyph designer, while using the rater's own knowledge and experience to estimate the learning time. Hopefully, there will be more empirical studies in the future to provide more data about the time required to learn different glyph designs in different contexts.

The rater needs to estimate the following values:
\begin{itemize}
    \item The \emph{learning time} required in the ``self-learning'' mode.
    \item The \emph{learning time} required in the ``self-learning + Q\&A'' mode, i.e., the potential users will likely need to hold a Q\&A session with the designer or someone who has already learned the encoding scheme of a glyph design.
    \item The \emph{learning time} required in the ``tutorial'' mode.
    \item The effort required for \emph{repeated learning}, in terms of four levels
    (i.e., effortless, minor effort, noticeable effort, and serious effort).
\end{itemize}

In summary, the rater can determine a Type D score by using the following steps:

\begin{enumerate}
    \item[1.] If the rater has the opportunity to meet potential users, observe the learnability of a glyph design in the meetings. Otherwise, gather the learnability information from the glyph designer who is expected to have the opportunity to meet potential users.
    \item[2.] Estimate how long it will take to learn the encoding scheme for all data variables in a glyph design in three learning modes (i.e., self learning, self-learning + Q\&A, and tutorial).
    \item[3.] Estimate the effort required for repeated learning in terms of four levels (i.e., effortless, minor effort, noticeable effort, and serious effort).
    \item[fin.] Map the three time values for different learning modes and the effort level for repeated learning to one of the five levels defined in Section \ref{sec:Learnability}.
\end{enumerate}


\subsection{Memorability}
\label{apx:Memorability}
Similar to learnability, an experience rater will likely be able to predict whether potential users can still remember a glyph design or part of a glyph design after 1 hour and 24 hours respectively. Of course, it is not absolutely necessary to define memorability based on the 1-hour and 24-hour interval. Nevertheless, it would be confusing and time consuming to consider many different intervals for evaluating this criterion.

A glyph designer or a rater can also use meetings to observe the memorability of a glyph design. For example, during a meeting, one may note down how often some participants would say ``I forgot what the black squire is for,'' or ask ``could you remind me what the middle line is?''. At the beginning of the subsequent meetings, a designer can also ask participants to recall the associations between data variables and visual channels.  

Similar to the evaluation of learnability, for an independent rater who may not have the access to the potential users, the rater can gather memorability information from the glyph designer, while using the rater's own knowledge and experience to estimate the level of easiness and difficulty in remembering a glyph design. Hopefully, there will be more empirical studies in the future to provide more data about the memorability of different glyph designs in different contexts.

In summary, the rater can determine a Type D score by using the following steps:

\begin{enumerate}
    \item[1.] If the rater has the opportunity to meet potential users, observe the memorability of a glyph design using the meetings. Otherwise, gather the memorability information from the glyph designer who is expected to have the opportunity to meet potential users.
    \item[2.] Scan all data variables in the glyph design, and estimate whether potential users can still remember, after 1 hour and 24 hours respectively, (i) the associations between the data variables and their visual channels and (ii) the encoding scheme of each visual channel in the glyph design, based on the information gathered, in conjunction with the rater's own analysis of the glyph design and knowledge and experience about glyph memorability in general.
    \item[3.] Work out the percentage of the data variables, whose visual encoding can still be remembered after 1 hour and after 24 hours.
    \item[fin.] Map the two percentage values to one of the five levels defined in Section \ref{sec:Memorability}.
\end{enumerate}



\end{document}